\documentclass[graybox]{svmult}

\usepackage{type1cm}        
\usepackage{makeidx}         
\usepackage{graphicx}        
\usepackage{multicol}        
\usepackage[bottom]{footmisc}

\usepackage{newtxtext}       %
\usepackage{newtxmath}       

\usepackage{url}

\makeindex             

\begin{document}

\title*{Meteors - light from comets and asteroids}
\author{Pavol Matlovi\v{c} and Juraj T\'{o}th}
\institute{Pavol Matlovi\v{c} \at Comenius University in Bratislava, \email{matlovic@fmph.uniba.sk}
\and Juraj T\'{o}th \at Comenius University in Bratislava, \email{toth@fmph.uniba.sk}}
%
%

\maketitle

\abstract{In studies of the oldest solar system bodies -- comets and asteroids -- it is their fragments -- meteoroids -- that provide the most accessible planetary material for detailed laboratory analysis in the form of dust particles or meteorites. Some asteroids and comets were visited by spacecrafts and returned interplanetary samples to Earth, while missions Hayabusa 2 and OSIRIX-REx visiting asteroids Ryugu and Bennu are ongoing. However, the lack of representative samples of comets and asteroids opens the space to gain more knowledge from direct observations of meteoroids. At collision with the Earth's atmosphere, meteoroids produce light phenomena known as meteors. Different methods can be used to observe meteors, allowing us to study small interplanetary fragments, which would otherwise remain undetected. Numerous impressive meteor showers, storms and meteorite impacts have occurred throughout the recorded history and can now be predicted and analyzed in much more detail. By understanding the dynamics, composition and physical properties of meteoroids, we are able to study the formation history and dynamical evolution of the solar system. This work presents an introduction to meteor astronomy, its fundamental processes and examples of current research topics.}

\section{Introduction: the space of meteoroids}
\label{sec:1}

\subsection{Overview}

Studying the solar system, as our home planetary neighborhood, the outlook for the first space travels, and the only recognized source of life has always been one of the main interests of astronomy and science in general. Despite the significant progress in our understanding of how the solar system was formed, what bodies constitute it, and what mechanisms influence their motion, there are still numerous unanswered questions regarding the complex nature of our planetary system. 

\begin{figure}[t]
\sidecaption[t]
\includegraphics[width=7.5cm]{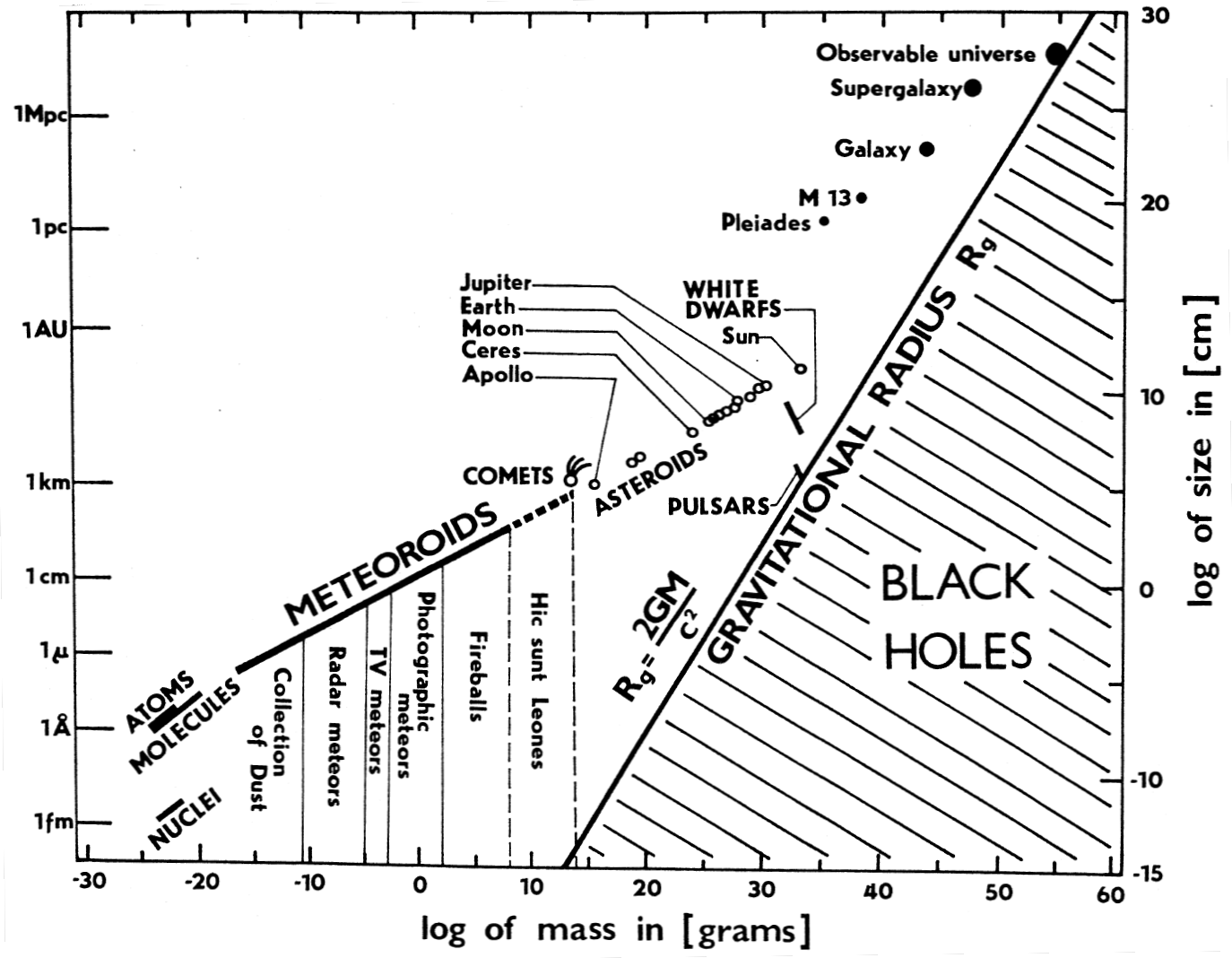}
\caption{Mass versus size diagram containing some known objects of the observable universe and showing the significance of the meteoroid complex (from \cite{1998SSRv...84..327C}).}
\label{fig:1}      
\end{figure}

Many of the key information we have about the solar system come from the studies of the oldest remnants left over from the planetary formation in the early protosolar disk - asteroids, comets, and meteoroids. On one side, these bodies can reveal the processes and conditions occurring in the early stages of the solar system. On the other, they can give rise to the dangers of catastrophic impacts, which have subjected our planet numerous times in the history. In each case, they are the topic of high scientific interest. The studies of asteroids and comets are however often complicated by the small size and low albedo of these bodies. Usually only limited information can be obtained from direct observations. Meteor observations during the interaction of meteoroids with the Earth's atmosphere allow us to study small solar system bodies, which would otherwise remain undetected. The importance of meteoroid studies was eloquently stressed by \cite{1998SSRv...84..327C}, by simply plotting the mass versus size diagram of objects in the observable universe (Fig. \ref{fig:1}). This plot demonstrates the multitude of meteoroids in the observable universe.

\begin{figure}[t]
\includegraphics[width=\textwidth]{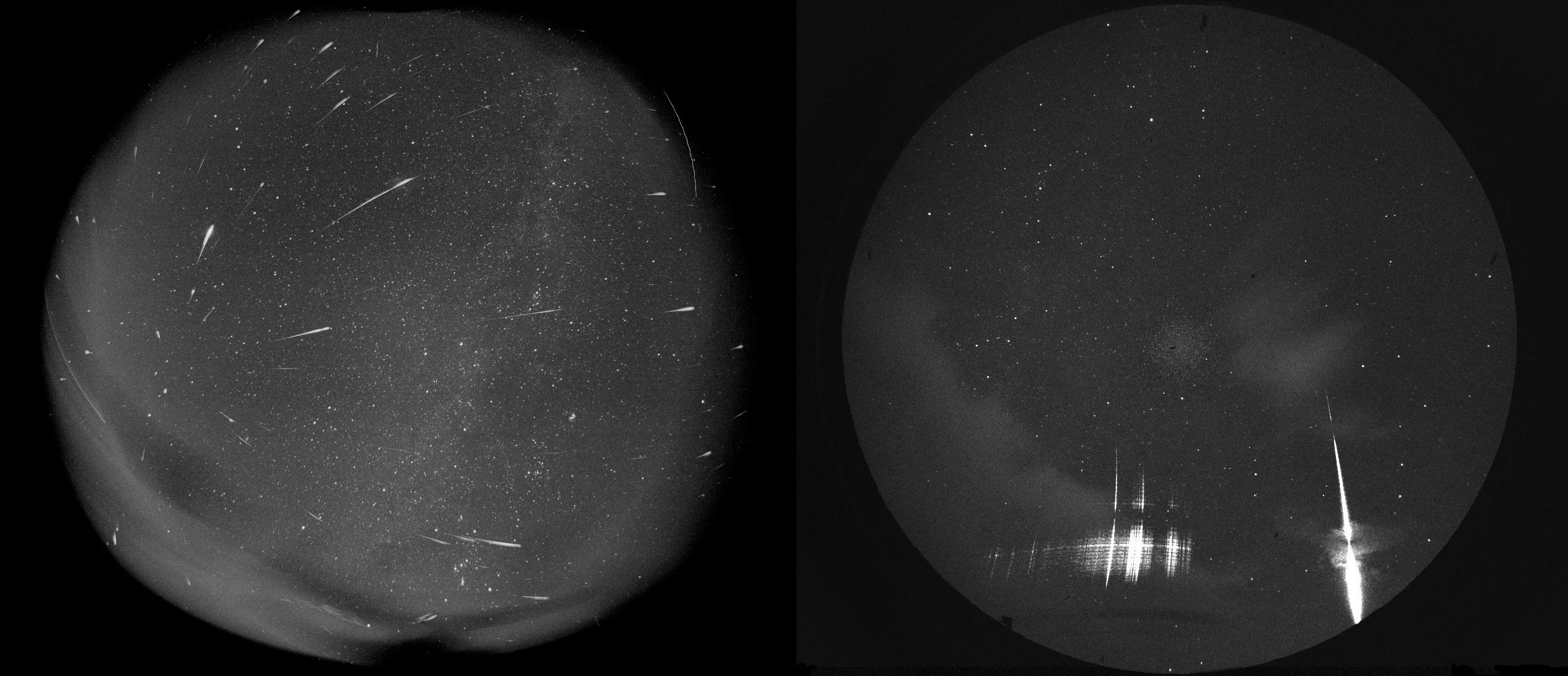}
\caption{Left: Leonid meteor shower observed during the 1998 outburst by a photographic camera at the Modra Observatory. Right: a fireball captured by All-sky Meteor Orbit System (AMOS) spectrograph, along with the first order emission spectrum (images by J. T\'{o}th and P. Matlovi\v{c}).}
\label{fig:2}  
\end{figure}

Recently, meteor astronomy has been gaining popularity among professional and amateur astronomers, due to the arising possibilities of using inexpensive technologies, particularly sensitive CCD cameras to effectively observe meteors and provide valuable scientific data. Most of these efforts are focused on studying the identification methods and activities of meteor showers \cite{2016Icar..266..331J, 2006mspc.book.....J}, determining original heliocentric meteoroid orbits or detecting potential meteorite impacts from the brightest fireballs \cite{2003Natur.423..151S, 2009Natur.458..485J, 2013Natur.503..238B, 2013M&PS...48.1757B}. Furthermore, the research focused on physical and compositional properties of meteoroids is progressing by applying photographic and video spectrographs to study emission spectra of meteors \cite{1993A&A...279..627B, 2003M&PS...38.1283T, 2005Icar..174...15B}. Physical parameters such us meteoroid masses, strengths and densities can be determined by studying their atmoshperic ablation and deceleration \cite{2004A&A...418..751C, 2007A&A...473..661B, 2011A&A...530A.113K}. Example of a meteor shower captured by a photographic system and a meteor spectrum observed by an all-sky video spectrograph is on Fig. \ref{fig:2}. 

The aim of this work is to provide basic introduction to meteor astronomy, its fundamental processes, observational techniques and relevant examples of specific research topics. A more comprehensive reviews of meteor studies beyond the presented individual examples can be found in \cite{1998SSRv...84..327C, 2019SSRv..215...34K, 2019msme.book.....R, 2018SSRv..214...23P, 2017P&SS..143..116J} and the references therein.

\subsection{Terms in meteor astronomy}
\label{sec:1.1}

\begin{figure}[t]
\sidecaption[t]
\includegraphics[width=7.5cm]{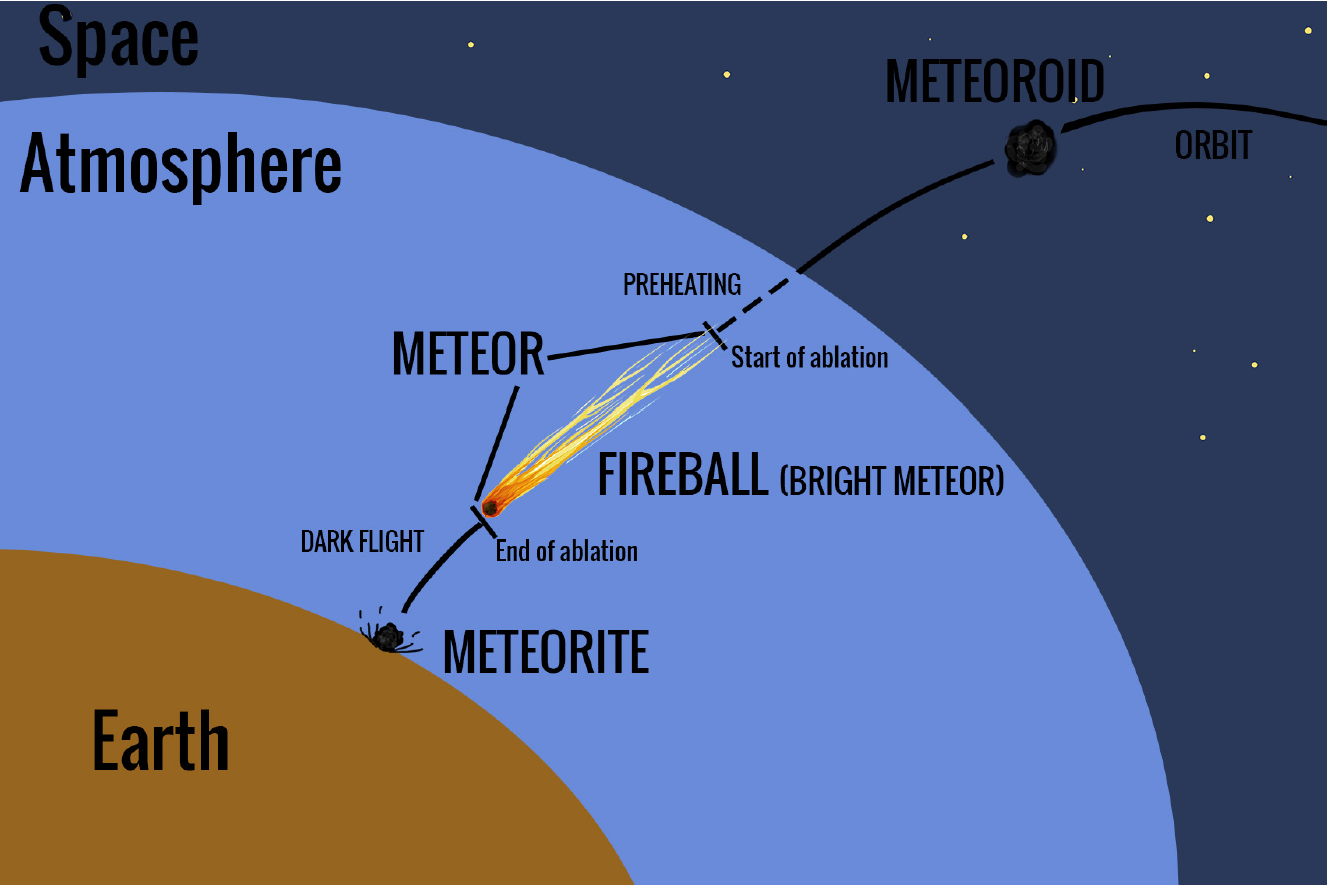}
\caption{Different phases of meteoroid interaction and characteristic terms in meteor astronomy (image by V. Voj\'{a}\v{c}ek).}
\label{fig:3}  
\end{figure}

\textit{Meteoroid} is currently defined as a solid natural object of a size roughly between 30 micrometers and 1 meter moving in, or coming from, interplanetary space\footnote{defined in 2017 by the IAU Commision F1 on Meteors, Meteorites and Interplanetary Dust}. The size limits have been set by agreement and do not represent a physical boundary. In the context of meteor observations, any object causing a meteor can be termed a meteoroid, irrespective of its size. Bodies smaller than 30 micrometers tend to radiate heat away more efficiently and not to vaporize during the atmospheric entry. These smaller bodies are known as \textit{interplanetary dust} particles. \textit{Meteorite} is any natural solid object that survived the meteor phase in a gaseous atmosphere without being completely vaporized. Meteorites smaller than 1 mm in size are also called micrometeorites. Depending on their speed, these may be too small to experience ablation in an atmosphere. Graphical interpretation of the meteor terminology is given in Fig. \ref{fig:3}.

Meteoroids are dominantly produced as the decay products of comets and asteroids. Only a minority of meteoroids come from the surfaces of planets (e.g. Mars) and planetary satellites (e.g. Moon) or from interstellar space. Though meteoroids are conglomerates of materials formed in primordial solar nebula, the dynamical lifetime of objects in near-Earth space is assumed to be of the order of 10 Myr \cite{2000Icar..146..176G,2000A&A...353..797F}. This means that no meteoroids could have stayed on current orbits from the beginning of the solar system.

We consider three main processes which lead to the separation of meteoroids from their parent bodies. Comets produce meteoroids through the process of sublimation and following gas drag \cite{1951ApJ...113..464W}. During the cometary activity near its perihelion, the drag of vapors from evaporating ices also releases solid particles, dust and meteoroids. Secondly, catastrophic disruption of comets can produce secondary nuclei and numerous dust particles and meteoroids (see Chapter 21 in \cite{2006mspc.book.....J}). The third process is related to direct collisions of solar system bodies, particularly among the main belt asteroids, which produce many collisional fragments \cite{2003ApJ...591..486N}. The separation velocities of these fragments are always much smaller than the original orbital velocity. Therefore, young meteoroids follow very similar orbits as their parent body. In this early stage, it is relatively easy to link the \textit{meteoroid stream} with its parent body. With time, gravitational perturbations from planets and various non-gravitational forces such as the Poynting-Robertson effect \cite{1979Icar...40....1B} cause the dispersion of the stream and separation from its parent orbit \cite{2006MNRAS.370.1841V}. 

In this respect we can distinguish stream meteoroids, which are usually just recently separated from their parent object (typically up to few thousand years) and \textit{sporadic meteoroids} on seemingly random orbits, significantly altered from the parent orbit. We assume that sporadic meteoroids were freed from their parent object thousands to millions of years ago. Compilation of known parent objects of meteoroid streams can be found e.g. in \cite{2007IAUS..236..107B} or \cite{2006mspc.book.....J}. The sporadic complex was described e.g. by \cite{2009Icar..201..295W, 2011ApJ...743..129N, 2018SSRv..214...64L}. The topics of meteor studies will be further discussed in Section \ref{sec:4}.
 									
Meteoroids may dynamically evolve to pass near Earth and interact with our atmosphere. The interaction may, for certain meteoroid sizes and velocities, result in a luminous phenomenon known as \textit{meteor}. Meteoroid streams entering the Earth's atmosphere generate so-called \textit{meteor showers}. For observers, meteor showers appear to originate from the same point and direction in the sky known as the \textit{radiant}. The name of a meteor shower is derived from the apparent position of the radiant in the night sky. For example, Leonid meteor shower has a radiant in the Leo constellation.

Owing to the interaction with air molecules, a meteoroid entering the Earth's atmoshpere heats up to high temperatures and starts to melt and vaporize. A column of ionized and excited plasma is produced along the meteoroid trail, producing light, ionization, and for larger particles, shock waves \cite{1983pmp..book.....B}. The meteor phenomenon can exhibit several phases. The brightest part is called the meteor \textit{head}. Ionization of air along the meteor path forms an ion \textit{train} which reflects radio waves in the decameter range. The ion trains of bright meteors can be visible even to the naked eye and those of particularly brilliant meteors may persist for seconds or even minutes (also known as persistent trains). \textit{Wake} of the meteor is the luminosity extending directly behind the meteor head and forming comet-like appearance of bright meteors. Meteor wake has different spectral features compared to the spectrum of meteor head. At a given position, the wake duration is only fraction of a second.

During the atmospheric flight, the meteoroid loses mass through processes of vaporization, fusion (melting), and fragmentation. Generally, the process of mass loss by a meteoroid is known as \textit{ablation}. The resistance of the atmosphere causes meteoroid to decelerate. Ablation and deceleration affect one another, since ablation depends on the meteoroid velocity and deceleration on its mass. Therefore, the equations describing deceleration and mass loss (see Eq. 3 and Eq. 5 in the next section) of the body must be solved simultaneously.

The collisions with air molecules and dynamical load generally cause the fragmentation of meteoroids. There are several ways in which meteoroids break up. The most significant fragmentation processes include progressive fragmentation \cite{1955AJ.....60Q.165J, 1969SSRv...10..230V}, in which the meteoroid fragments into parts which continue to crumble; (quasi-)continuous fragmentation \cite{1975MNRAS.173..339H, 2002A&A...384..317B} describing continuous detachment of small particles; and gross (sudden) fragmentation \cite{1993A&A...279..615C, 2007A&A...473..661B} characteristic during brilliant bursts, in which meteoroid suddenly disrupts into large number of fragments. 

Recent studies suggest that most meteoroids undergo some form of fragmentation during the meteor phase. The most successfully applied ablation and fragmentation models follow the concept of the dustball meteoroid \cite{2004A&A...418..751C, 2007A&A...473..661B}. Still, many of the methods we use to study meteoroid properties assume meteoroid interaction as a single non-fragmenting body (next Section). While this approximation is sufficient for most of our estimates, precise description of meteor deceleration and mass loss must be based on a model of effective fragmentation. Neglecting meteoroid fragmentation was one of the main reasons behind the discrepancies of determined masses in past analyses (e.g. the discussion in \cite{1967SCoA...11...35C, 1967SCoA...11...61V}.

\section{Atmospheric interaction: basics of meteor physics}
\label{sec:2}

The motion and ablation of a meteoroid in the Earth's atmosphere is most often described by the single body theory (Chapter 3 in \cite{1998SSRv...84..327C}). Single body theory refers to the mass loss, deceleration, luminosity and ionization related to the motion of a single non-fragmenting body. The theory assumes that the heat transfer, ionization, luminosity and drag coefficient are during this path constant \cite{1983pmp..book.....B}. The presented standard equations of single body theory (following the formalization used by \cite{2012Weryk}) are used to describe meteors both before and after their disruption.

Collisions of atmospheric molecules with meteoroid can either liberate atoms from the surface of a meteoroid directly (also known as sputtering), or heat the material to its boiling point of approximately 2000 K. At these temperatures, material starts to evaporate from the surface of the meteoroid in a phenomenon known as thermal ablation. It is assumed that thermal ablation is the dominant process of mass loss (see \cite{2005P&SS...53.1341R} and the discussion therin). Excess of thermal stress or stagnation pressure can cause meteoroid to fragment into numerous smaller pieces. These fragments continue to collide with atmospheric molecules and ablate on their own as single bodies. The ablated material colliding with atmospheric molecules produces a trail of ionized and excited plasma. Rather than individual atomic emission, ablation may also take the form of a dust emission/removal, releasing a trail of heated small particles which cause the meteor wake \cite{1998SSRv...84..327C}. The ablation behavior differs significantly among meteoroids. Large variations are observed in the beginning heights and light curve shapes among smaller meteoroids, and in end heights among larger meteoroids.

The basis of the mathematical form describing the motion of a meteoroid in atmosphere was first presented by \cite{1938PAPS...79..499W} who used Hoppe's solution with constant coefficients. More elaborate solutions to these differential solutions were later presented by \cite{1956BAICz...7...58L} and \cite{1983pmp..book.....B}.

Let us assume meteoroid passing through a distance $v \Delta$t in time period $\Delta t$. A meteoroid with a cross-section area $S$ will encounter atmospheric mass (with atmospheric density $\rho_a$) of $m_a = \rho_a S v\Delta t$. The cross-section area can be rewritten by introducing the dimensionless shape factor $A = S/V^{2/3}$. Meteoroid volume V is related to meteoroid bulk density and meteoroid mass as $V = m/ \rho_m$. We will assume that $A$ is constant and for simplicity usually corresponds to a sphere (A=1.21). The rate of air mass encountering the meteoroid is defined as:

\begin{equation}
\frac{dm_a}{dt} = \frac{\Delta{}m_a}{\Delta{}t}=\frac{A v {\rho{}}_a m^{\frac{2}{3}}}{{{\rho{}}_m}^{\frac{2}{3}}}
\end{equation}

\noindent
We can express the momentum transfer from the atmosphere to the meteoroid by:

\begin{equation}
\frac{d\left(mv\right)}{dt} = \frac{dm}{dt}v+\frac{dv}{dt}m=\Gamma{}v\frac{dm_a}{dt}
\end{equation}

The drag coefficient $\Gamma$ in Eq. 2 is defined as the fraction of momentum transferred to the body from the ongoing molecules of air. The drag coefficient can vary between 0 (for no transfer of momentum) and 2 (perfect reflection of air molecules). For small meteoroids, the term $dm/dt$ can be neglected \cite{1998SSRv...84..327C}. Next, by substituting Eq. 1 into Eq. 2, we get the drag equation:

\begin{equation}
\frac{dv}{dt} = -\frac{\Gamma{}A{\rho{}}_av^2}{{\rho{}}_m^{\frac{2}{3}}m^{\frac{1}{3}}}
\end{equation}

\noindent
The drag equation is the first fundamental equation in meteor physics. It describes the deceleration of a meteoroid during its flight in the atmosphere (the deceleration is emphasized by the negative sign at the right-hand side of the equation).

The second fundamental equation is called the mass-loss equation. The mass loss rate is determined by the kinetic energy transferred from the atmosphere to the meteoroid. We assume that a certain fraction of the kinetic energy of the oncoming air molecules is expended on ablation of mass (vaporization or fusion and spraying) of the meteoroid \cite{1983pmp..book.....B}. The loss of mass during ablation can be expressed as:

\begin{equation}
\frac{dm}{dt} = -\frac{\Lambda{}E_a}{\xi{}\Delta{}t} = -\frac{\Lambda{}v^2}{2\xi{}}\frac{dm_a}{dt}
\end{equation}

\noindent
where $E_a$ is the kinetic energy of interacting air molecules and $\xi$ is the heat of ablation, representing the energy required to melt/vaporize one unit of meteoroid mass $dm$. Substituting Eq. 1 into Eq. 4 gives the conventional form of the mass-loss equation:

\begin{equation}
\frac{dm}{dt} = -\frac{\Lambda{}A{\rho{}}_av^3m^{\frac{2}{3}}}{2\xi{}{\rho{}}_m^{\frac{2}{3}}}
\end{equation}

The heat-transfer coefficient $\Lambda$ is valued between zero and unity, since the energy used on ablation cannot exceed the total kinetic energy of interacting air molecules. Part of this kinetic energy will be used to heat up the body of a meteoroid, part will be re-radiated, and part will be expended for excitation and ionization of the meteoroid and surrounding air molecules. If fragmentation takes place, some of this energy is also expended to break the mechanical bonds between meteoroid grains. 		

It is assumed that the amount of light produced during this process is also related to the mass-loss rate, as it is proportional to the kinetic energy lost by the meteoroid \cite{1998SSRv...84..327C}. The energy released by the meteoroid in the form of radiation, typically in the visible spectrum is described in the luminosity equation: 

\begin{equation}
I=-\tau{}\frac{dE_m}{dt}=-\tau{}\left(\frac{v^2}{2}\frac{dm}{dt}+\frac{dv}{dt}mv\right)
\end{equation}

\noindent
Here, $I$ is the radiative power (bolometric or in specific band pass) and $\tau$ is the luminous efficiency, which is defined as the fraction of the kinetic energy loss of a meteoroid transformed into radiation. Generally, the luminous efficiency is dependent on the wavelength of radiation, the chemical composition of the meteoroid body and the atmosphere, on the meteoroid velocity and possibly on the meteoroid mass. The deceleration term $(dv/dt)$ can be neglected for fast and faint meteors \cite{1998SSRv...84..327C}.

Most of the meteor radiation comes from line emissions in evaporated meteoroid atoms \cite{1983pmp..book.....B}. Clearly, the chemical composition of the meteoroid plays significant role in the nature of the produced emission, since different chemical elements are represented by different line strengths in the visible spectrum. The ionization produced during the interaction with atmosphere can be described using the ionization equation:

\begin{equation}
q=-\frac{\beta{}}{\mu{}v}\frac{dm}{dt}
\end{equation}

\noindent
In this equation, $q$ is the electron line density, which represents the number of electrons per unit trail length. The ionization coefficient $\beta$ describes the average number of electrons produced per ablated atom, while the atomic mass of standard meteoroid atom is labeled as $\mu$. The amount of ionization is again dependent on the mass-loss rate.

To determine meteoroid mass from the observed photon and electron count, we need the values of  $\tau$ and $\beta$, which is often problematic. Usual process of obtaining the ionization mass is based on measuring the electron line density $q$, assuming and ionization curve and integrating Eq. 7. Similarly, the photometric mass is determined by measuring $I$ along the meteor trail and integrating Eq. 6.

Equations 3, 5, 6 and 7 contain several variables that need to be determined in order to solve the fundamental equations of meteor physics. Some of them are given by the physical properties of the meteoroid, some are determined from observations and others can be estimated theoretically. The atmospheric density $\rho_a$ is usually defined by a model of the atmosphere for different heights of meteoroid-atmosphere interaction. Currently the most commonly used video observations allow us to determine the velocity of a meteor as a function of time and luminosity. The heat of ablation is defined by the character of ablation (vaporizing or spraying) and is thus dependent on the composition of a meteoroid. Lastly, the luminous efficiency, heat transfer coefficient and drag coefficient must be obtained from theoretical models, experiments, or by estimating based on observational data.

\section{The nature of meteor radiation: spectra and meteoroid composition}
\label{sec:3}

Meteor radiation is produced mainly by the excitation of atoms, due to the mutual collisions with atmospheric atoms and molecules, and due to the recombination of free electrons in the surrounding ionized gas and subsequent cascade transitions \cite{1983pmp..book.....B}. The radiation of a meteor originates in the plasma envelope of air and meteoric vapor surrounding the meteoroid. Meteor spectra consist primarily of atomic emission lines and molecular bands. Spectral analyses indicate that it is mainly the atoms and ions of the meteoroid vapor which radiate. Although meteor spectra have been observed since 1864, the knowledge gained from these observations is rather scarce. Most early studies were focused on the description of the spectra and identification of lines. The most extensive identifications are given by \cite{1961PDOO..25..3C} for Perseid meteors (velocity of app. 60 km\,s\textsuperscript{-1}), by \cite{1971BAICz..22..219C} for a 32 km\,s\textsuperscript{-1} meteor and by \cite{1994A&AS..103...83B} for a 19 km\,s\textsuperscript{-1} meteor. Overview of reliably identified atoms and ions is presented e.g. in Section 3.3 of \cite{1998SSRv...84..327C}.		

Not all lines in meteor spectra can be explained by a single temperature. \cite{1994P&SS...42..145B} revealed that meteor spectra are composed of two distinct components with different characteristic temperatures. The lower temperature component is called the main spectrum and its origin is in the radiating gas of meteoroid and atmospheric vapors. The temperature lies usually in the range 3500 - 5000 K and does not generally depend on meteor velocity. The main spectrum consists of several hundreds of lines, mostly neutral lines of atoms of meteoric origin. The most notable lines present in the main spectrum are of Fe I, Mg I, Na I, Ca I, Cr I, Mn I and Ca II. Thermal equilibrium is nearly satisfied, although some lines may deviate. It was shown that chemical composition of the radiating plasma can be computed from the main spectrum. 		

The high temperature component is also known as the second spectrum and has characteristic temperature of nearly 10000 K. The high temperature region is probably formed in the front of the meteoroid, near the shock wave of the meteor \cite{1993A&A...279..627B}. The chemical composition cannot be derived exactly, nevertheless, the determined elemental abundances from fireball spectra were found to be consistent with common meteorite composition. The typical lines for the second spectrum are the high excitation lines of Mg II, Si II, N I and O I. The low excitation transitions in singly ionized atoms can be present in both spectral components. This is most notably the case of Ca II, which is bright in both spectra, but also fainter lines of Ti II and Sr II can be present in both components. The second spectrum is strong in fast meteors while it can be absent in slow meteors with velocity of about 15 km\,s\textsuperscript{-1} \cite{1998SSRv...84..327C}. The ratio of gas masses involved in the production of both components was found to be a steep function of velocity.

The ratio of meteoric vapors to the atmospheric species in both components shows interesting disparities. Nitrogen and oxygen lines are not present in the main component. This is caused by the absence of allowed low excitation transitions in these atoms. The atoms simply do not radiate at 5000 K. Based on the pressure balance with surrounding atmosphere, \cite{1993A&A...279..627B} concluded that about 95\% of atoms in low temperature gas were the invisible atmospheric species. The second spectrum demonstrates both meteoritic and atmospheric emissions. Their ratio varies widely. In faint meteors of medium and high velocity the meteoric emissions in the second spectrum are often absent, while O I and N I lines and N\textsubscript{2} bands are still present. The N\textsubscript{2} bands sometimes appear very early on the trajectory \cite{1973NASSP.319..153C}. Other works \cite{1995EM&P...71..237B} described cases in which meteoritic emission invisible at the start of the trajectory burst out later, while the atmospheric lines brighten only moderately. Based on these effects, it appears that atmospheric emissions are less dependent on the ablation rate, which is rather expected.		

The previously described spectral lines are characteristic for meteor head - the brightest part of a meteor. Other phases of the meteor phenomenon present specific spectral features. The spectrum of the meteor wake consists chiefly from low excitation lines. Typical wake lines belong to Na I, Fe I, Mg I and Ca I. The short-duration trains are formed by only one spectral line, the forbidden green auroral line of neutral atomic oxygen at 557.7 nm. The luminosity is probably produced by the atmospheric oxygen. Persistent trains are still not well understood phenomena. Several spectra have been taken in the recent years, which show different features from case to case. The spectra show both continuous or quasi-continuous radiation and atomic lines. The most important and most persistent line, common for all spectra, is the sodium doublet near 589.2 nm. This suggests that the long-living luminosity is due to similar mechanism which produces the sodium airglow - a luminous layer in the Earth's mesosphere (80 - 105 km) of characteristic yellow color.

While meteor spectroscopy presents the most efficient way to study of meteoroid composition from unbiased sources, the composition of the radiating plasma during meteoric interaction does not fully reflects the composition of the original meteoroid \cite{1993A&A...279..627B}. Meteoroid composition must cover a wide range of material types including asteroidal, lunar and martian samples known from meteorite laboratory studies (for mineralogical overview, see \cite{papike1998planetary}) and the lesser known cometary materials, so far only covered by few spacecraft probes such as the Stardust mission to comet 81P/Wild 2 \cite{2012M&PS...47..453B} and Rosetta mission to comet 67P/Churyumov-Gerasimenko \cite{2016Icar..271...76L}.

\begin{figure}[t]
\sidecaption[t]
\includegraphics[width=7.5cm]{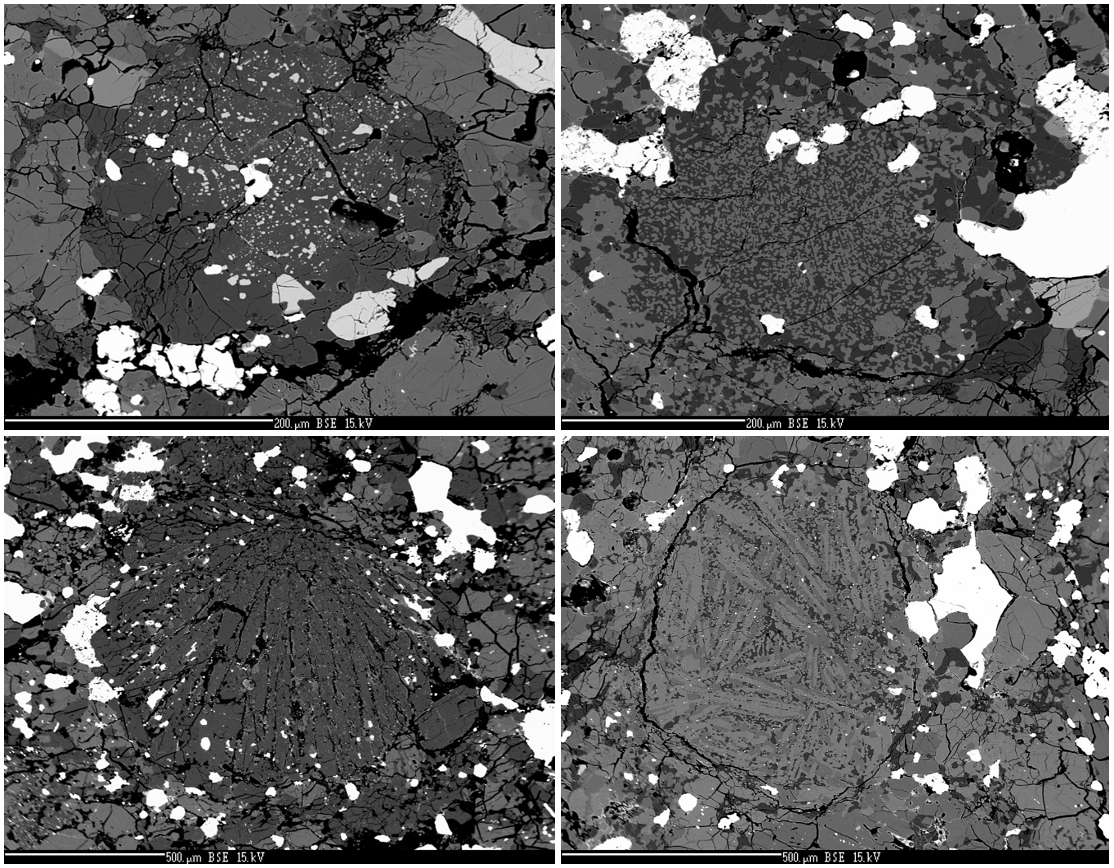}
\caption{Different types of chondrules (large circular shapes in the middle of the pictures) in the Ko\v{s}ice ordinary chondrite meteorite (image by D. Ozdin, further description is provided in \cite{2015M&PS...50..864O}).}
\label{fig:4}  
\end{figure}

\begin{figure}[t]
\sidecaption[t]
\includegraphics[width=7.5cm]{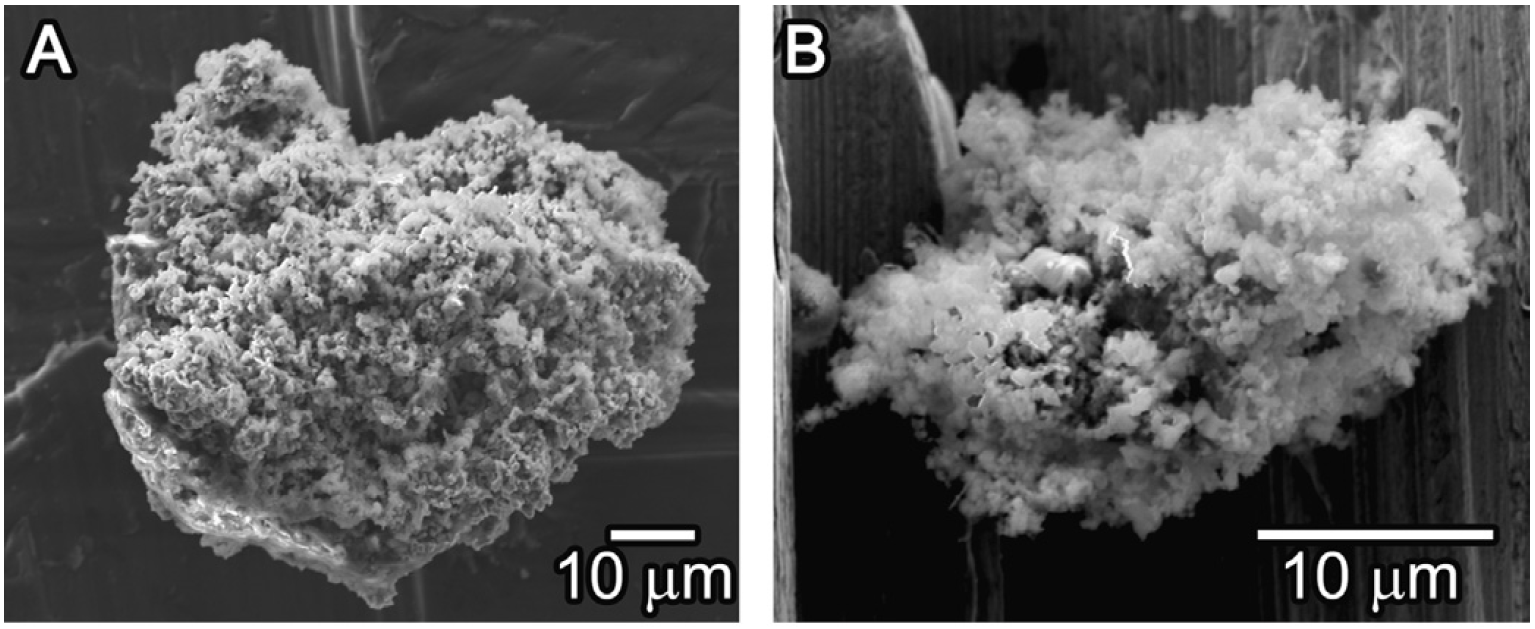}
\caption{Electron images of chondritic porous micrometeorites of cometary dust (from \cite{2015E&PSL.410....1N}).}
\label{fig:5}  
\end{figure}

Most of the known meteorite samples come from primitive rocky asteroidal bodies formed in the early solar system which never reached the melting limit temperature in their interiors. This primitive material presents valuable chemical clues preserved from the early stages of the solar system formation. Most of the known rocky meteorites are chondritic, meaning that they contain chondrules, 0.1 - 1-mm sized objects of glass and crystalline silicates formed by melting followed by rapid cooling (Fig. \ref{fig:4}). They also contain Ca- and Al-rich inclusions (CAIs) composed of refractory oxides and silicates and formed by condensation of high-temperature nebular gases. Besides the primordial material from the early solar system, it was detected that chondrites also contain particles from other stars \cite{2003Sci...300..105M}. Primitive chondritic composition has been also detected in samples of cometary dust (Fig. \ref{fig:5}).

Not all interplanetary bodies are chondritic. Variations of stony achondrites, iron and stony-iron materials have been identified from meteorite samples on Earth. They point towards more complicated formation processes and material differentiation in larger bodies. Still, together they only constitute approximately 7\% of all known meteorite samples. The ratios of identified meteorite materials is however biased by the fact that only stronger asteroidal samples can withstand the ablation in the atmosphere and reach the Earth's surface. Given the dust production rates of comets and the observed activities of major meteor showers, it is likely that majority of smaller meteoroids in interplanetary space are of cometary origin. Precise orbital data determined from multi-station meteor observations along with the compositional information from emission spectroscopy allow us to study the real distribution of materials in the solar system.

\begin{figure}[!tbp]
  \centering
  \begin{minipage}[b]{0.49\textwidth}
    \includegraphics[width=\textwidth]{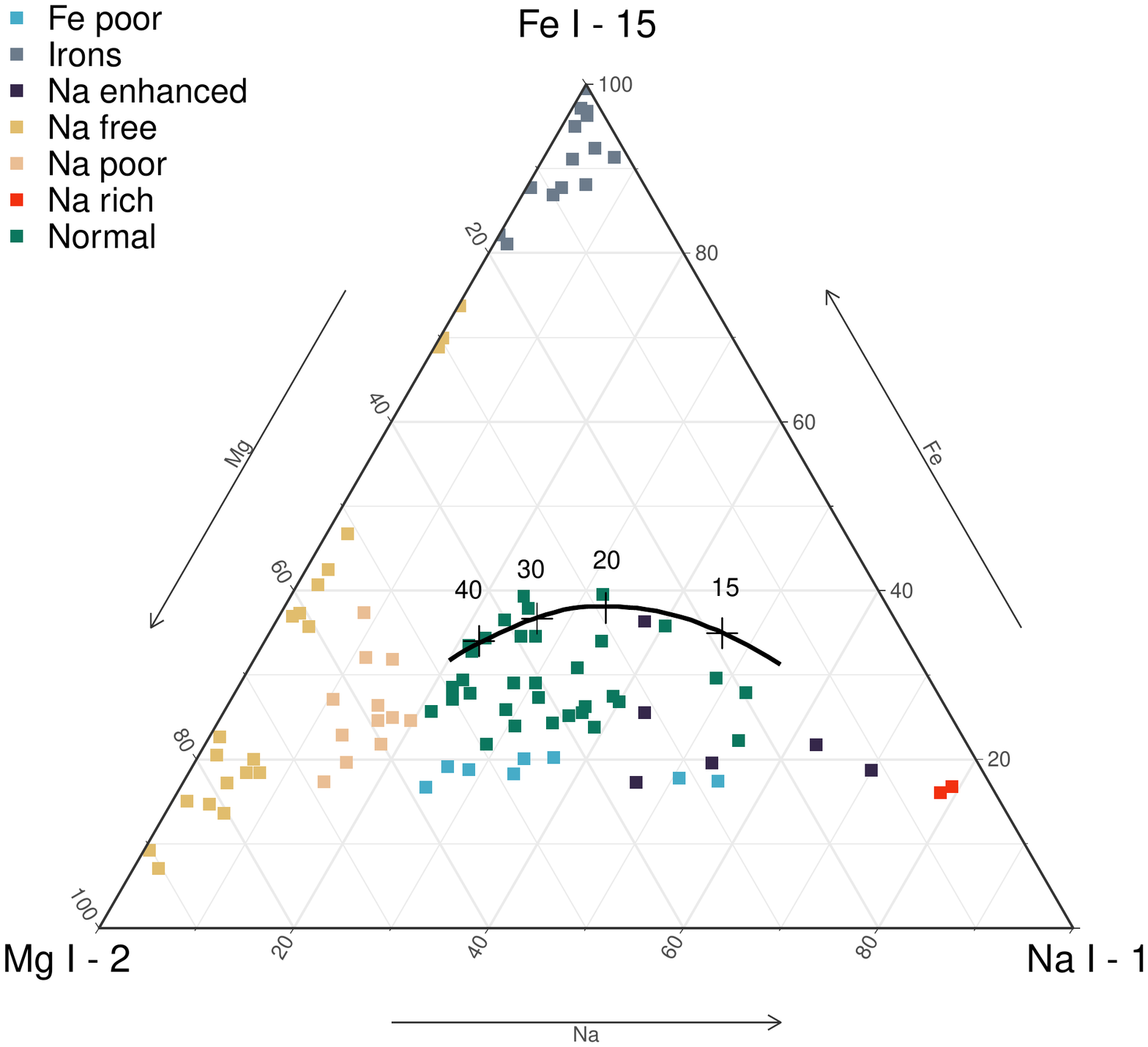}
  \end{minipage}
  \hfill
  \begin{minipage}[b]{0.49\textwidth}
    \includegraphics[width=\textwidth]{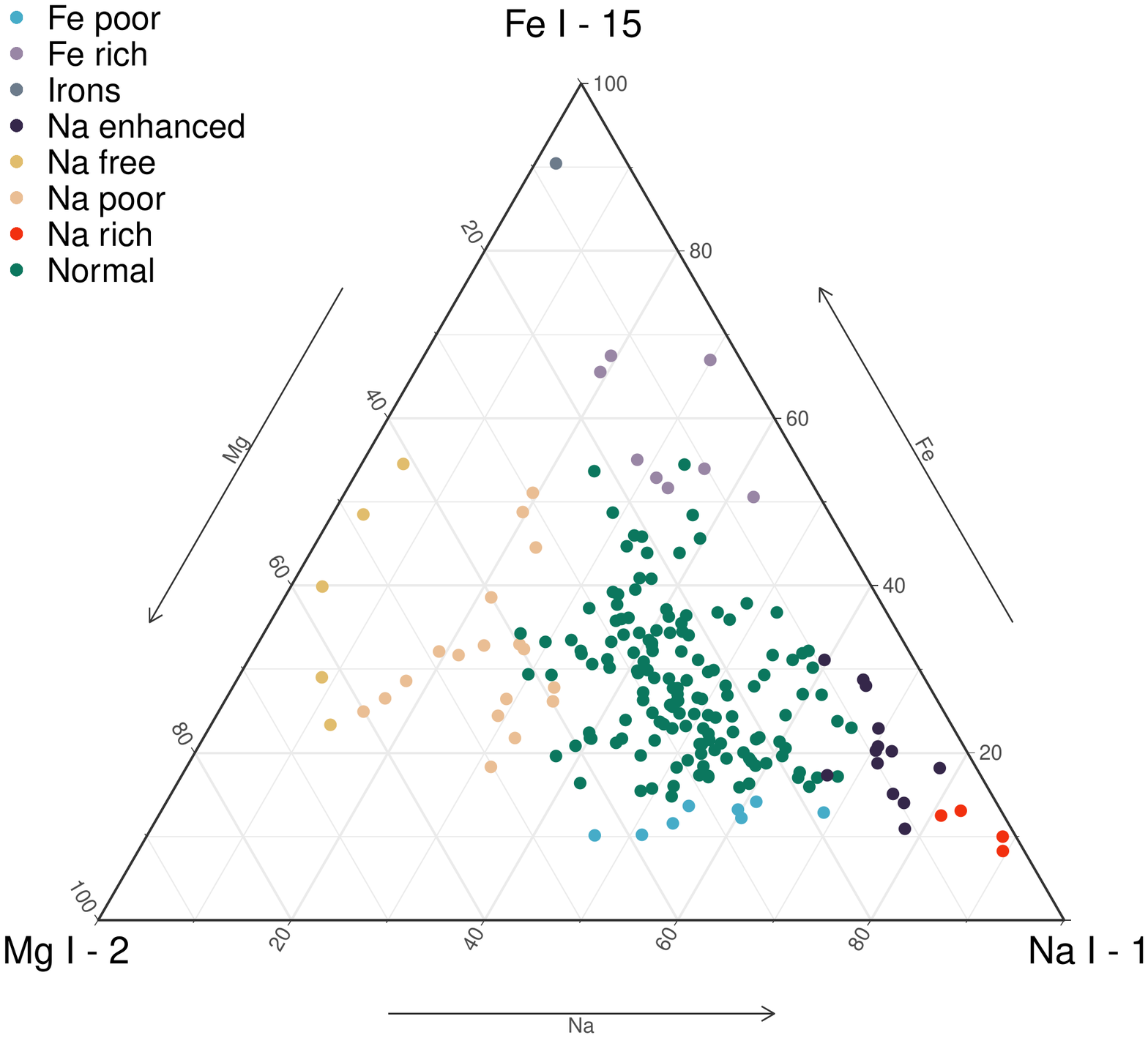}
  \end{minipage}
   \caption{Spectral classification of mm-sized and mm-m-sized meteoroids showing range of material types detected from meteor observations (from \cite{2005Icar..174...15B} and \cite{2019A&A...629A..71M}).}
   \label{fig:6}
\end{figure}

To achieve this goal, larger surveys of meteor spectra based sensitive video spectrographs were initiated \cite{2005Icar..174...15B, 2015A&A...580A..67V, 2016P&SS..123...25R}. Generally, these instruments yield low-resolution spectra from which elemental abundances cannot be reliably determined. To reveal variations of meteoroid composition, the method of spectral classification was established. The method is based on the relative intensities of the three main emission multiplets Na I, Mg I and Fe I representative of the different components of meteoroid composition (volatile, silicate and metallic respectively). Fig. \ref{fig:6} shows the results of such studies for different size populations of meteoroids.

While this method can be used to distinguish distinct meteoroid types from common chondritic bodies, the distinction between specific material types (e.g. specific chondritic classes) is difficult \cite{2015aste.book..257B}. For this purpose, detailed model of meteor spectrum based on the solution of radiative transfer needs to be applied to records captured in high-resolution. First such model assuming thermal equilibrium and self-absorption in the radiating plasma was developed by \cite{1993A&A...279..627B} and applied to an excellent spectrum captured by a photographic system. Relative abundances for 9 elements were determined (Fe, Na, Mg, Ca, Ti, Cr, Mn, Ni, and Al) in agreement with laboratory meteorite measurements. Similar method has been since used for individual fireball spectra by few different authors \cite{2003M&PS...38.1283T, 2018A&A...613A..54D}. Similarly detailed analyses of meteor spectra for larger quantities of meteoroids samples are still missing.

\begin{figure}[t]
\sidecaption[t]
\includegraphics[width=7.5cm]{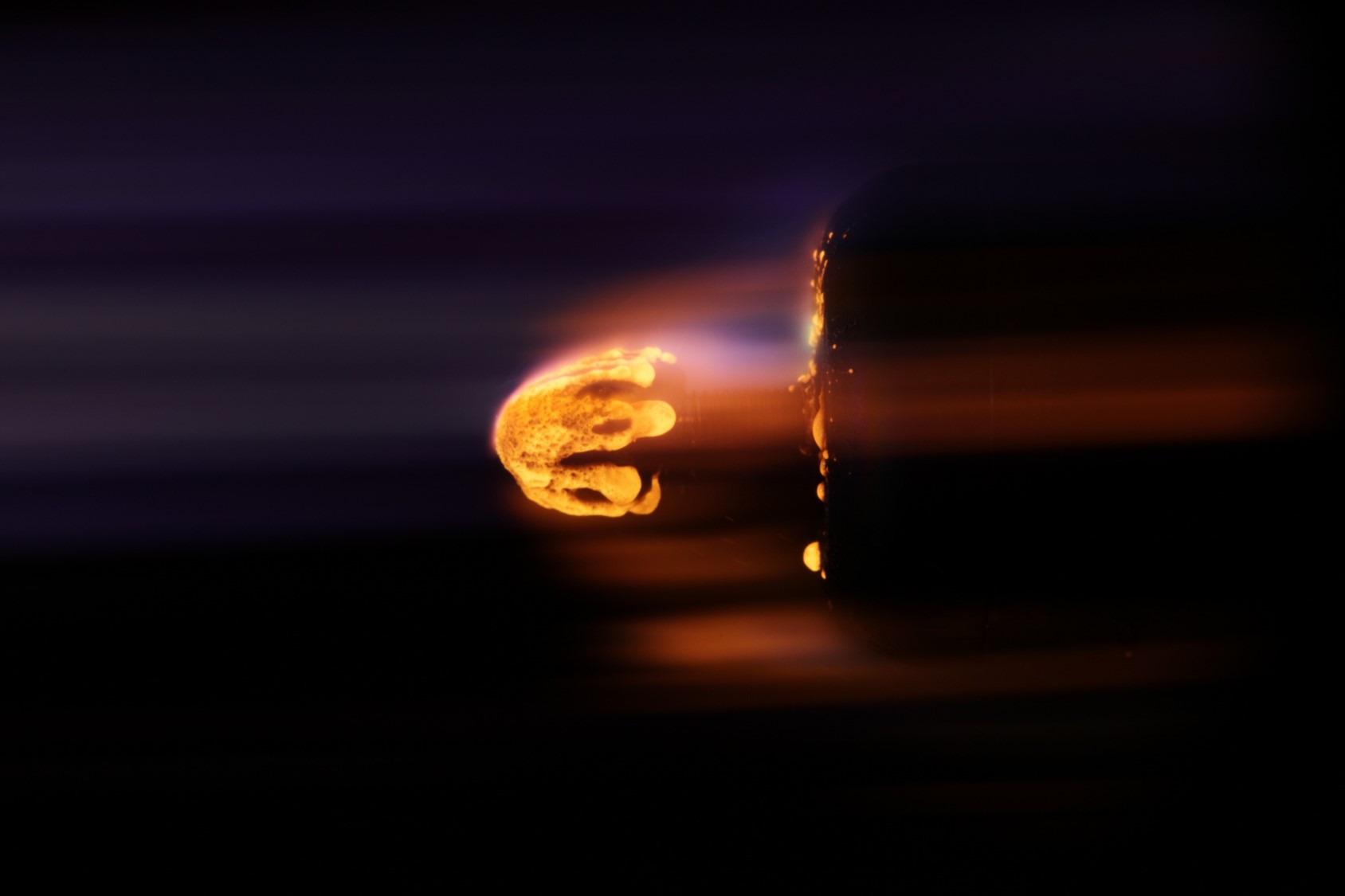}
\caption{The simulated ablation of a chondritic meteorite at the probe of the plasma wind tunnel (image by the High Enthalpy Flow Diagnostics Group, IRS).}
\label{fig:7}  
\end{figure}

Recent experiments suggest that our abilities to study meteoroid composition and ablation from ground-based observations can be also improved by laboratory analyses. Several teams have focused on using laser-induced breakdown spectroscopy of meteorite samples to study meteoroid composition \cite{2018A&A...610A..73F, 2018AcSpe.147...87D}. Alternatively, the simulated ablation of meteorites in plasma wind tunnels has been used to successfully reproduce the atmospheric meteoroid interaction \cite{2017ApJ...837..112L, 2018A&A...613A..54D, 2019ApJ...876..120H} (Fig. \ref{fig:7}). Besides high-resolution Echelle spectra, these experiments provide valuable data on the ablation processes for different meteoroid types. The next step is to quantitatively link the phenomena observed in the laboratory with real meteor observations in the atmosphere.

\section{Meteor observations and meteoroid population studies}
\label{sec:4}

For meteor scientists, Earth's atmosphere serves as a large detector of meteoroid inflow in the region of 1 au from the Sun. Various observational methods can be used to observe meteors and provide different information about meteoroids. The light from meteor ablation can be observed visually by naked eye or telescopes, by photographic and video cameras, radar systems, seismic and infrasonic detectors.

Simple visual observations of meteors, though the least comprehensive, have historically been crucial in gaining basic knowledge of the inflow of meteors and describing shower activities. The earliest records of individual meteor sightings and probable meteor showers were made in ancient Mesopotamia and China, at the beginning of second millennium BC \cite{1958SCoA....2..131I, 1973mman.book.....B}. Still today, visual reports are useful for constraining information on potential meteorite impacts, or calibration of instrumental efficiencies. Reports from visual observations can be submitted to the International Meteor Organization database\footnote{http://www.imo.net}.

Optical video and photographic observations are currently the most utilized by astronomers, due to their ability to precisely determine meteor fluxes and constrain atmospheric trajectories and speeds of meteoroids. By triangulation of meteor trajectories observed from multiple stations and backwards propagation of the directional and velocity information, original heliocentric orbit of a meteoroid can be determined (e.g. \cite{1987BAICz..38..222C}). Numerous groups have initiated development of video or photographic networks to provide large sky coverage with multi-station observations. These networks include the European Fireball Network \cite{1998M&PS...33...49O} in Central Europe, Cameras for Allsky Meteor Surveillance (CAMS) network \cite{2011Icar..216...40J} in Northern America, Desert Fireball Network in Australia \cite{2012AuJES..59..177B}, global All-sky Meteor Orbit System (AMOS) network \cite{2015P&SS..118..102T} based in Slovakia, Fireball Recovery and InterPlanetary Observation Network (FRIPON) in France \cite{2015pimo.conf...37C}, and many others. Orbital data from these networks is collected in databases such as the SonotaCo \cite{2010JIMO...38...54V}, EDMOND \cite{2014me13.conf..225K} and CAMS database \cite{2016Icar..266..331J}.

Trajectories and speeds of meteors can be also determined by radar systems operating in the 15 to 500 MHz frequency range. Radars can be used to study inflow of even the faintest meteors (caused by roughly micrometer particles) and can operate even during the daytime, allowing to identify activities of daytime showers. Some of the most renowned radar surveys providing crucial meteor data include the Advanced Meteor Orbit Radar Facility (AMOR) in New Zealand \cite{1994QJRAS..35..293B}, the Canadian Meteor Orbit Radar (CMOR) \cite{2004EM&P...95..221W} or the Southern Argentina Agile MEteor Radar (SAAMER) \cite{2015ApJ...809...36J}. Radar observations have been also used to constrain the mass-dependent flux of interplanetary particles on Earth \cite{2016A&A...592A.150P, 2011MNRAS.414.3322B}. The standard mass distribution was previously determined from lunar crater counts and taking into account meteoroid and interplanetary dust measurements \cite{1985Icar...62..244G}.  

During observations, meteor showers caused by Earth intersecting meteoroid streams appear to originate from the same point and direction in the sky, known as the radiant. These meteoroids were ejected from their parent comet or asteroid relatively recently and can be used to probe the properties of larger parent bodies (e.g. \cite{2006CoSka..36..103P, 1993MNRAS.264...93A, 2017A&A...605A..68S, 2010AJ....139.1822J} or \cite{2006mspc.book.....J} for overview). For some major streams, the association to parent object is well known (e.g. Perseids and comet 109P/Swift-Tuttle, Orionids and comet 1P/Halley or Draconids and 21P/Giacobini-Zinner). However, there are numerous minor streams and newly detected streams (for overview, see the IAU Meteor Data Center\footnote{https://www.ta3.sk/IAUC22DB/MDC2007/} \cite{2017P&SS..143....3J}), which are yet to be confirmed in activity and linked to their parent object. A new review of the current state of minor streams a sporadic background research is presented in \cite{2019msme.book..210W}.

In some cases, the link to a meteoroid stream or a parent body can be established based on orbital similarity. Dissimilarity parameters such as the D-criterion based on orbital \cite{1963SCoA....7..261S, 1981Icar...45..545D} or geocentric parameters \cite{1999MNRAS.304..743V, 2008EM&P..102...73J} can be used. Recent works focus on applications or improvements of these methods to identify potential meteoroid associations in orbital databases \cite{2016MNRAS.455.4329M, 2017A&A...598A..40N, 2015P&SS..118...38R}. Alternatively, wavelet transform methods can be used to search for meteoroid streams in radar data \cite{2008Icar..195..317B}. Furthermore, modeling of the meteoroid stream evolution and prediction of meteor shower activities has improved significantly with the development of more powerful computational facilities \cite{2005A&A...439..761V, 2013SoSyR..47..219R}.

Sporadic meteors are not associated with any meteoroid streams, but they dominate the meteoroid influx at Earth. They include particles from the interplanetary meteoroid cloud (from previous asteroid and comet dust production processes) and in small proportion also interstellar particles \cite{2018SSRv..214...64L}. The identification of their origin is hampered by orbital alteration caused by gravitational, radiation and collision processes. While among brighter meteors, sporadic particles comprise comparable datasets as shower meteors, their proportion increases dramatically among fainter meteors corresponding to smaller meteoroids (e.g. 99\% of the meteors in the AMOR radar database \cite{2005MNRAS.359..551G}). This effect is caused by different ejections velocities and radiation forces affecting the evolution of smaller particles. All of the instrumentally observed meteorite impacts have originated from sporadic meteors \cite{2015aste.book..257B}. The improved observational techniques and larger sky coverage allows for higher efficiency in the location of meteorite impacts \cite{2003Natur.423..151S, 2009Natur.458..485J, 2013Natur.503..238B, 2013M&PS...48.1757B}. The most advanced techniques for trajectory and photometric measurements today allow for very precise prediction of the meteorite strewn field \cite{2017M&PS...52.1683B}.

Large datasets of sporadic meteoroids from optical and radar detections have revealed six sporadic meteor sources at the Earth. These sources do not generally correspond to physical meteoroid structures, but rather describe observed concentrated regions in the radiant space. The detected meteor sources are affected by the motion of the Earth around the Sun \cite{1993MNRAS.265..524J}. The strongest sources are the helion and anti-helion source in the ecliptic plane at aproximately 70$^{\circ}$ from the apex direction. These sources are likely formed by particles produced by Jupiter-family comets \cite{2009Icar..201..295W, 2010ApJ...713..816N}. We also observe north and south apex sources centered towards the direction of Earth's movement and toroidal sources 60$^{\circ}$ north and south from the apex. The apex and toroidal sources seem to originate from Halley-type or long-period comets \cite{2009Icar..201..295W, 2014ApJ...789...25P}. Furthermore, radiant distribution of sporadic meteors reveals a ring depleted in meteor radiants at 55$^{\circ}$ from the apex \cite{2008Icar..196..144C}. The ring is attributed to high-inclunation meteoroids undergoing Kozai oscillations \cite{2009Icar..201..295W}. Studies of sporadic meteors have also provided constrains for the speed distribution of incoming particles, showing that most meteoroids impact the Earth at low speeds ($\approx$ 11--20 km\,s\textsuperscript{-1}) \cite{1999P&SS...47.1005M, 2017P&SS..143..209M}. 

Meteor observations can be also used to study the presence of interstellar bodies in the solar system. Interstellar dust was first detected by the Ulysses spacecraft \cite{1993Natur.362..428G} during a flyby of Jupiter and since studied by multiple other missions. The measured interstellar particles are usually not larger than a few microns. The detection of larger interstellar meteoroids would provide significant implications for the dust-to-gas mass ratio in close interstellar medium. Generally, meteoroids with speed above approximately 72 km\,s\textsuperscript{-1} (given by the sum of the escape speed from the solar system and orbital speed of Earth) are on orbits not bound to the Sun. Slower meteoroids can still originate from the interstellar space, but cannot be identified as such only based on their speed. Typically large uncertainties of the determined meteoroid velocities for very fast meteors must be taken into account. The first detection of interstellar meteoroids was reported from the AMOR radar data \cite{1996Natur.380..323T}, though these results are still being debated. Analysis of CMOR radar observations showed that the interstellar meteoroids would present a fraction too small to be statistically meaningful \cite{2004EM&P...95..221W}. A search in the video meteor databases has revealed that majority of identified hyperbolic meteoroids are caused by the velocity estimation error \cite{2014M&PS...49...63H}. 

As briefly discussed in previous section, observations of meteor emission spectra can provide information on meteoroid composition. Generally, these studies are either focused on detailed analyses of exceptional fireballs and can provide relative element abundances \cite{1993A&A...279..627B, 2003M&PS...38.1283T, 2007AdSpR..39..491J}, focus on characterization of meteoroid streams \cite{2017P&SS..143..104M, 2014EM&P..112...45R} and their parent bodies, or utilize larger datasets of low-resolution spectra to study variations of meteoroid composition from different orbital sources in the solar system \cite{2005Icar..174...15B, 2015A&A...580A..67V, 2019A&A...629A..71M, 2017P&SS..143..238M}. Meteor spectra studies have revealed that the depletion of volatiles in meteoroids is mainly caused by solar radiation in close proximity of the Sun and partially by cosmic-ray irradiation in the Oort cloud \cite{2005Icar..174...15B}. Depletion of sodium was also shown to have implications for meteoroid structure and material strength \cite{2019A&A...621A..68V}. Besides atomic emission lines, several molecular bands have been identified in fireball spectra \cite{2016Icar..278..248B}. One of the interesting goals of meteor spectroscopy is to detect organic matter in meteoroids. While the detection of organic carbon can be difficult in meteor spectra \cite{2004AsBio...4...67J, 2010Icar..210..150B}, the commonly observed hydrogen H\textsubscript{$\alpha$} line can be used as a tracer for the presence of organics and water \cite{2004AsBio...4..123J}. 

Meteor trajectories and their light curves can be used to study physical properties of meteoroids. Meteoroid masses can be estimated based on optical meteor brightness or electron line density in radar data. Unfortunately, the accuracy of mass estimation is still limited mainly by the uncertainties of luminous and ionization efficiency parameters \cite{2012JGRA..117.9323C}. The most consistent estimates of meteoroid masses were yielded by models combining meteor light curves and atmospheric deceleration \cite{2007A&A...473..661B, 2011Icar..212..877G}. The beginning and terminal heights of meteor luminous trajectory can be used to infer the material strength of a meteoroid. The empirical classification of \cite{1988BAICz..39..221C} differentiates between the most fragile cometary (Draconid-type), standard and dense cometary, carbonaceous and ordinary chondritic material strengths. Using the meteoroid dustball model \cite{2004A&A...418..751C}, distribution of bulk densities of meteoroids was studied by \cite{2011A&A...530A.113K}. It was revealed that on average, meteoroids on asteroidal orbits have densities of 4200 kg\,m\textsuperscript{-3}, on Jupiter-family orbits 3100 kg\,m\textsuperscript{-3} and between 260 and 1900 kg\,m\textsuperscript{-3} on Halley-type orbits. Grain densities of meteoroids can also be estimated using the heat conductivity equation and combining meteor trajectory measurements with laboratory data for different rock and mineral types \cite[and references therein]{2009A&A...495..353B}. The difference between the bulk and grain density can be used to infer porosities of meteoroids.

Recent works and high-definition meteor observations \cite{2018DPS....5010001V} suggest that majority of meteoroids undergo different forms of fragmentation in the atmosphere, which is not accounted for in the single body theory (Section \ref{sec:2}). Furthermore, it has been shown that ablation of meteoroids composed of iron-nickel alloy differs significantly from standard chondritic meteoroids \cite{2019A&A...625A.106C}. For accurate determination of meteoroid physical properties, more complex models that describe the ablation and fragmentation of meteoroids are required \cite{2007A&A...473..661B, 2013A&A...557A..41C, 2015MNRAS.447.1580S}. 

\section{Summary}
\label{sec:5}

We have presented a brief introduction to meteor astronomy and the fundamental processes of meteor ablation. The approach to constrain meteoroid properties from various ground-based observation techniques was described. Meteoroids are small interplanetary bodies continuously produced by sublimation, outgassing and collisions of comets and asteroids. During their interaction with the Earth's atmosphere, they create luminous phenomena known as meteors. This process can be studied from various points of view and eventually allows us to study dynamical processes and distribution of interplanetary material in the solar system. The current meteor research topics focus on different aspects of the phenomenon. Due to improved and efficient video and photographic techniques, numerous observational networks were created around the world and produce large datasets of orbital data. These observations can be used to study activities of meteor showers and link meteoroid streams with parent comets and asteroids. 

Due to the improved sky coverage, the number of instrumentally observed meteorite falls rises and allows higher efficiency in locating meteorite impacts. The sensitive radar observations enable studies of interplanetary dust inflow, sources of the sporadic background and activities of daytime showers. More complex models of meteoroid ablation and emission spectra can be used to determine physical properties and composition of meteoroids and provide implication for the processes of material transfer in the solar system. Future analyses of meteors can be improved by utilizing laboratory facilities such us wind tunnels and shock tubes to simulate meteor ablation in controlled environment. While we already know a lot about the processes of meteor interaction and meteoroid populations, there are still numerous open questions to be answered and more details be obtained to better understand our solar system and prepare for potential Earth impacts and spacecraft shielding. 
 
\begin{acknowledgement}
This work was supported by the ERASMUS+ project 2017-1-CZ01-KA203-035562, by the Slovak Research and Development Agency grant APVV-16-0148 and the Slovak Grant Agency for Science grant VEGA 01/0596/18.
\end{acknowledgement}

\bibliographystyle{spphys}
\bibliography{references}

\begin{thebibliography}{100}
\providecommand{\url}[1]{{#1}}
\providecommand{\urlprefix}{URL }
\expandafter\ifx\csname urlstyle\endcsname\relax
  \providecommand{\doi}[1]{DOI \discretionary{}{}{}#1}\else
  \providecommand{\doi}{DOI \discretionary{}{}{}\begingroup
  \urlstyle{rm}\Url}\fi

\bibitem{1998SSRv...84..327C}
Z.~{Ceplecha}, J.~{Borovi{\v c}ka}, W.G. {Elford}, D.O. {Revelle}, R.L.
  {Hawkes}, V.~{Porub{\v c}an}, M.~{{\v S}imek}, Space Science Reviews
  \textbf{84}, 327 (1998).
\newblock \doi{10.1023/A:1005069928850}

\bibitem{2016Icar..266..331J}
P.~{Jenniskens}, Q.~{N{\'e}non}, J.~{Albers}, P.S. {Gural}, B.~{Haberman},
  D.~{Holman}, R.~{Morales}, B.J. {Grigsby}, D.~{Samuels}, C.~{Johannink},
  Icarus \textbf{266}, 331 (2016).
\newblock \doi{10.1016/j.icarus.2015.09.013}

\bibitem{2006mspc.book.....J}
P.~{Jenniskens}, \emph{{Meteor Showers and their Parent Comets}} (2006)

\bibitem{2003Natur.423..151S}
P.~{Spurn{\'y}}, J.~{Oberst}, D.~{Heinlein}, Nature \textbf{423}, 151 (2003).
\newblock \doi{10.1038/nature01592}

\bibitem{2009Natur.458..485J}
P.~{Jenniskens}, M.H. {Shaddad}, D.~{Numan}, S.~{Elsir}, A.M. {Kudoda}, M.E.
  {Zolensky}, L.~{Le}, G.A. {Robinson}, J.M. {Friedrich}, D.~{Rumble}, Nature
  \textbf{458}(7237), 485 (2009).
\newblock \doi{10.1038/nature07920}

\bibitem{2013Natur.503..238B}
P.G. {Brown}, J.D. {Assink}, L.~{Astiz}, R.~{Blaauw}, M.B. {Boslough},
  J.~{Borovi{\v{c}}ka}, N.~{Brachet}, D.~{Brown}, M.~{Campbell-Brown},
  L.~{Ceranna}, Nature \textbf{503}(7475), 238 (2013).
\newblock \doi{10.1038/nature12741}

\bibitem{2013M&PS...48.1757B}
J.~{Borovi{\v c}ka}, J.~{T{\'o}th}, A.~{Igaz}, P.~{Spurn{\'y}}, P.~{Kalenda},
  J.~{Haloda}, J.~{Svore{\aa}}, L.~{Korno{\v s}}, E.~{Silber}, P.~{Brown},
  M.~{Hus{\'a}Rik}, Meteoritics and Planetary Science \textbf{48}, 1757 (2013).
\newblock \doi{10.1111/maps.12078}

\bibitem{1993A&A...279..627B}
J.~{Borovi{\v c}ka}, Astronomy \& Astrophysics \textbf{279}, 627 (1993)

\bibitem{2003M&PS...38.1283T}
J.M. {Trigo-Rodriguez}, J.~{Llorca}, J.~{Borovi{\v c}ka}, J.~{Fabregat},
  Meteoritics and Planetary Science \textbf{38}, 1283 (2003).
\newblock \doi{10.1111/j.1945-5100.2003.tb00313.x}

\bibitem{2005Icar..174...15B}
J.~{Borovi{\v c}ka}, P.~{Koten}, P.~{Spurn{\'y}}, J.~{Bo{\v c}ek}, R.~{{\v
  S}tork}, Icarus \textbf{174}, 15 (2005).
\newblock \doi{10.1016/j.icarus.2004.09.011}

\bibitem{2004A&A...418..751C}
M.D. {Campbell-Brown}, D.~{Koschny}, Astronomy \& Astrophysics \textbf{418},
  751 (2004).
\newblock \doi{10.1051/0004-6361:20041001-1}

\bibitem{2007A&A...473..661B}
J.~{Borovi{\v c}ka}, P.~{Spurn{\'y}}, P.~{Koten}, Astronomy \& Astrophysics
  \textbf{473}, 661 (2007).
\newblock \doi{10.1051/0004-6361:20078131}

\bibitem{2011A&A...530A.113K}
J.B. {Kikwaya}, M.~{Campbell-Brown}, P.G. {Brown}, Astronomy \& Astrophysics
  \textbf{530}, A113 (2011).
\newblock \doi{10.1051/0004-6361/201116431}

\bibitem{2019SSRv..215...34K}
D.~{Koschny}, R.H. {Soja}, C.~{Engrand}, G.J. {Flynn}, J.~{Lasue}, A.C.
  {Levasseur-Regourd}, D.~{Malaspina}, T.~{Nakamura}, A.R. {Poppe}, V.J.
  {Sterken}, Space Science Reviews \textbf{215}(4), 34 (2019).
\newblock \doi{10.1007/s11214-019-0597-7}

\bibitem{2019msme.book.....R}
G.O. {Ryabova}, D.J. {Asher}, M.J. {Campbell-Brown}, \emph{{Meteoroids: Sources
  of Meteors on Earth and Beyond}} (2019)

\bibitem{2018SSRv..214...23P}
J.M.C. {Plane}, G.J. {Flynn}, A.~{M{\"a}{\"a}tt{\"a}nen}, J.E. {Moores}, A.R.
  {Poppe}, J.D. {Carrillo-Sanchez}, C.~{Listowski}, Space Science Reviews
  \textbf{214}(1), 23 (2018).
\newblock \doi{10.1007/s11214-017-0458-1}

\bibitem{2017P&SS..143..116J}
P.~{Jenniskens}, Planetary and Space Science \textbf{143}, 116 (2017).
\newblock \doi{10.1016/j.pss.2017.01.008}

\bibitem{2000Icar..146..176G}
B.~{Gladman}, P.~{Michel}, C.~{Froeschl{\'e}}, Icarus \textbf{146}, 176 (2000).
\newblock \doi{10.1006/icar.2000.6391}

\bibitem{2000A&A...353..797F}
L.~{Foschini}, P.~{Farinella}, C.~{Froeschl{\'e}}, R.~{Gonczi}, T.J. {Jopek},
  P.~{Michel}, Astronomy \& Astrophysics \textbf{353}, 797 (2000)

\bibitem{1951ApJ...113..464W}
F.L. {Whipple}, The Astrophysical Journal \textbf{113}, 464 (1951).
\newblock \doi{10.1086/145416}

\bibitem{2003ApJ...591..486N}
D.~{Nesvorn{\'y}}, W.F. {Bottke}, H.F. {Levison}, L.~{Dones}, The Astrophysical
  Journal \textbf{591}, 486 (2003).
\newblock \doi{10.1086/374807}

\bibitem{1979Icar...40....1B}
J.A. {Burns}, P.L. {Lamy}, S.~{Soter}, Icarus \textbf{40}, 1 (1979).
\newblock \doi{10.1016/0019-1035(79)90050-2}

\bibitem{2006MNRAS.370.1841V}
J.~{Vaubaillon}, P.~{Lamy}, L.~{Jorda}, Monthly Notices of the Royal
  Astronomical Society \textbf{370}, 1841 (2006).
\newblock \doi{10.1111/j.1365-2966.2006.10606.x}

\bibitem{2007IAUS..236..107B}
J.~{Borovi{\v c}ka}, in \emph{Near Earth Objects, our Celestial Neighbors:
  Opportunity and Risk}, \emph{IAU Symposium}, vol. 236, ed. by G.B.
  {Valsecchi}, D.~{Vokrouhlick{\'y}}, A.~{Milani} (2007), \emph{IAU Symposium},
  vol. 236, pp. 107--120.
\newblock \doi{10.1017/S1743921307003134}

\bibitem{2009Icar..201..295W}
P.~{Wiegert}, J.~{Vaubaillon}, M.~{Campbell-Brown}, Icarus \textbf{201}, 295
  (2009).
\newblock \doi{10.1016/j.icarus.2008.12.030}

\bibitem{2011ApJ...743..129N}
D.~{Nesvorn{\'y}}, D.~{Janches}, D.~{Vokrouhlick{\'y}}, P.~{Pokorn{\'y}}, W.F.
  {Bottke}, P.~{Jenniskens}, The Astrophysical Journal \textbf{743}(2), 129
  (2011).
\newblock \doi{10.1088/0004-637X/743/2/129}

\bibitem{2018SSRv..214...64L}
A.C. {Levasseur-Regourd}, J.~{Agarwal}, H.~{Cottin}, C.~{Engrand}, G.~{Flynn},
  M.~{Fulle}, T.~{Gombosi}, Y.~{Langevin}, J.~{Lasue}, T.~{Mannel}, Space
  Science Reviews \textbf{214}(3), 64 (2018).
\newblock \doi{10.1007/s11214-018-0496-3}

\bibitem{1983pmp..book.....B}
V.A. {Bronshten}, \emph{{Physics of meteoric phenomena}} (1983)

\bibitem{1955AJ.....60Q.165J}
L.~{Jacchia}, Astronomical Journal \textbf{60}, 165 (1955).
\newblock \doi{10.1086/107133}

\bibitem{1969SSRv...10..230V}
F.~{Verniani}, Space Science Reviews \textbf{10}, 230 (1969).
\newblock \doi{10.1007/BF00212686}

\bibitem{1975MNRAS.173..339H}
R.L. {Hawkes}, J.~{Jones}, Monthly Notices of the Royal Astronomical Society
  \textbf{173}, 339 (1975).
\newblock \doi{10.1093/mnras/173.2.339}

\bibitem{2002A&A...384..317B}
P.B. {Babadzhanov}, Astronomy \& Astrophysics \textbf{384}, 317 (2002).
\newblock \doi{10.1051/0004-6361:20020010}

\bibitem{1993A&A...279..615C}
Z.~{Ceplecha}, P.~{Spurny}, J.~{Borovi{\v c}ka}, J.~{Keclikova}, Astronomy \&
  Astrophysics \textbf{279}, 615 (1993)

\bibitem{1967SCoA...11...35C}
Z.~{Ceplecha}, Smithsonian Contributions to Astrophysics \textbf{11}, 35 (1967)

\bibitem{1967SCoA...11...61V}
F.~{Verniani}, Smithsonian Contributions to Astrophysics \textbf{11}, 61 (1967)

\bibitem{2012Weryk}
R.~{Weryk}, {Simultaneous radar and video meteors}.
\newblock Ph.D. thesis, The University of Western Ontario ~(Canada) (2012)

\bibitem{2005P&SS...53.1341R}
L.A. {Rogers}, K.A. {Hill}, R.L. {Hawkes}, Planetary and Space Science
  \textbf{53}(13), 1341 (2005).
\newblock \doi{10.1016/j.pss.2005.07.002}

\bibitem{1938PAPS...79..499W}
F.L. {Whipple}, Proceedings of the American Philosophical Society \textbf{79},
  499 (1938)

\bibitem{1956BAICz...7...58L}
B.J. {Levin}, Bulletin of the Astronomical Institutes of Czechoslovakia
  \textbf{7}, 58 (1956)

\bibitem{1961PDOO..25..3C}
I.~{Halliday}, Publ. Dominion Obs. Ottawa \textbf{25}, 3 (1961)

\bibitem{1971BAICz..22..219C}
Z.~{Ceplecha}, Bulletin of the Astronomical Institutes of Czechoslovakia
  \textbf{22}, 219 (1971)

\bibitem{1994A&AS..103...83B}
J.~{Borovi{\v c}ka}, Astronomy \& Astrophysics \textbf{103}, 83 (1994)

\bibitem{1994P&SS...42..145B}
J.~{Borovi{\v c}ka}, Planetary and Space Science \textbf{42}, 145 (1994).
\newblock \doi{10.1016/0032-0633(94)90025-6}

\bibitem{1973NASSP.319..153C}
A.F. {Cook}, C.L. {Hemenway}, P.M. {Millman}, A.~{Swider}, NASA Special
  Publication \textbf{319}, 153 (1973)

\bibitem{1995EM&P...71..237B}
J.~{Borovi{\v c}ka}, J.~{Bo{\v c}ek}, Earth Moon and Planets \textbf{71}, 237
  (1995).
\newblock \doi{10.1007/BF00612965}

\bibitem{papike1998planetary}
J.J. Papike, \emph{Planetary materials}, vol.~36 (Mineralogical Society of
  America, 1998)

\bibitem{2012M&PS...47..453B}
D.~{Brownlee}, D.~{Joswiak}, G.~{Matrajt}, Meteoritics and Planetary Science
  \textbf{47}(4), 453 (2012).
\newblock \doi{10.1111/j.1945-5100.2012.01339.x}

\bibitem{2016Icar..271...76L}
Y.~{Langevin}, M.~{Hilchenbach}, N.~{Ligier}, S.~{Merouane}, K.~{Hornung},
  C.~{Engrand}, R.~{Schulz}, J.~{Kissel}, J.~{Ryn{\"o}}, P.~{Eng}, Icarus
  \textbf{271}, 76 (2016).
\newblock \doi{10.1016/j.icarus.2016.01.027}

\bibitem{2015M&PS...50..864O}
D.~{Ozd{\'{\i}}n}, J.~{Plav{\v c}an}, M.~{Hor{\aa}{\'a}{\v c}kov{\'a}},
  P.~{Uher}, V.~{Porub{\v c}an}, P.~{Veis}, J.~{Rakovsk{\'y}}, J.~{T{\'o}th},
  P.~{Kone{\v c}n{\'y}}, J.~{Svore{\aa}}, Meteoritics and Planetary Science
  \textbf{50}, 864 (2015).
\newblock \doi{10.1111/maps.12405}

\bibitem{2015E&PSL.410....1N}
T.~{Noguchi}, N.~{Ohashi}, S.~{Tsujimoto}, T.~{Mitsunari}, J.P. {Bradley},
  T.~{Nakamura}, S.~{Toh}, T.~{Stephan}, N.~{Iwata}, N.~{Imae}, Earth and
  Planetary Science Letters \textbf{410}, 1 (2015).
\newblock \doi{10.1016/j.epsl.2014.11.012}

\bibitem{2003Sci...300..105M}
S.~{Messenger}, L.P. {Keller}, F.J. {Stadermann}, R.M. {Walker}, E.~{Zinner},
  Science \textbf{300}(5616), 105 (2003).
\newblock \doi{10.1126/science.1080576}

\bibitem{2015A&A...580A..67V}
V.~{Voj{\'a}{\v c}ek}, J.~{Borovi{\v c}ka}, P.~{Koten}, P.~{Spurn{\'y}},
  R.~{{\v S}tork}, Astronomy \& Astrophysics \textbf{580}, A67 (2015).
\newblock \doi{10.1051/0004-6361/201425047}

\bibitem{2016P&SS..123...25R}
R.~{Rudawska}, J.~{T{\'o}th}, D.~{Kalman{\v c}ok}, P.~{Zigo}, P.~{Matlovi{\v
  c}}, Planetary and Space Science \textbf{123}, 25 (2016).
\newblock \doi{10.1016/j.pss.2015.11.018}

\bibitem{2019A&A...629A..71M}
P.~{Matlovi{\v{c}}}, J.~{T{\'o}th}, R.~{Rudawska}, L.~{Korno{\v{s}}},
  A.~{Pisar{\v{c}}{\'\i}kov{\'a}}, Astronomy \& Astrophysics \textbf{629}, A71
  (2019).
\newblock \doi{10.1051/0004-6361/201936093}

\bibitem{2015aste.book..257B}
J.~{Borovi{\v c}ka}, P.~{Spurn{\'y}}, P.~{Brown}, \emph{{Small Near-Earth
  Asteroids as a Source of Meteorites}} (2015), pp. 257--280.
\newblock \doi{10.2458/azu_uapress_9780816532131-ch014}

\bibitem{2018A&A...613A..54D}
A.~{Drouard}, P.~{Vernazza}, S.~{Loehle}, J.~{Gattacceca}, J.~{Vaubaillon},
  B.~{Zanda}, M.~{Birlan}, S.~{Bouley}, F.~{Colas}, M.~{Eberhart},
  T.~{Hermann}, L.~{Jorda}, C.~{Marmo}, A.~{Meindl}, R.~{Oefele},
  F.~{Zamkotsian}, F.~{Zander}, Astronomy \& Astrophysics \textbf{613}, A54
  (2018).
\newblock \doi{10.1051/0004-6361/201732225}

\bibitem{2018A&A...610A..73F}
M.~{Ferus}, J.~{Koukal}, L.~{Len{\v z}a}, J.~{Srba}, P.~{Kubel{\'{\i}}k},
  V.~{Laitl}, E.M. {Zanozina}, P.~{V{\'a}{\v n}a}, T.~{Kaiserov{\'a}},
  A.~{Kn{\'{\i}}{\v z}ek}, P.~{Rimmer}, E.~{Chatzitheodoridis}, S.~{Civi{\v
  s}}, Astronomy \& Astrophysics \textbf{610}, A73 (2018).
\newblock \doi{10.1051/0004-6361/201629950}

\bibitem{2018AcSpe.147...87D}
M.~{Dell'Aglio}, M.~{L{\'o}pez-Claros}, J.J. {Laserna}, S.~{Longo}, A.~{De
  Giacomo}, Spectrochimica Acta \textbf{147}, 87 (2018).
\newblock \doi{10.1016/j.sab.2018.05.024}

\bibitem{2017ApJ...837..112L}
S.~{Loehle}, F.~{Zander}, T.~{Hermann}, M.~{Eberhart}, A.~{Meindl},
  R.~{Oefele}, J.~{Vaubaillon}, F.~{Colas}, P.~{Vernazza}, A.~{Drouard},
  J.~{Gattacceca}, The Astrophysical Journal \textbf{837}, 112 (2017).
\newblock \doi{10.3847/1538-4357/aa5cb5}

\bibitem{2019ApJ...876..120H}
B.~{Helber}, B.~{Dias}, F.~{Bariselli}, L.F. {Zavalan}, L.~{Pittarello},
  S.~{Goderis}, B.~{Soens}, S.J. {McKibbin}, P.~{Claeys}, T.E. {Magin}, The
  Astrophysical Journal \textbf{876}(2), 120 (2019).
\newblock \doi{10.3847/1538-4357/ab16f0}

\bibitem{1958SCoA....2..131I}
S.~{Imoto}, I.~{Hasegawa}, Smithsonian Contributions to Astrophysics
  \textbf{2}, 131 (1958)

\bibitem{1973mman.book.....B}
J.K. {Bjorkman}, \emph{{Meteors and meteorites in the ancient Near East}}
  (1973)

\bibitem{1987BAICz..38..222C}
Z.~{Ceplecha}, Bulletin of the Astronomical Institutes of Czechoslovakia
  \textbf{38}, 222 (1987)

\bibitem{1998M&PS...33...49O}
J.~{Oberst}, S.~{Molau}, D.~{Heinlein}, C.~{Gritzner}, M.~{Schindler},
  P.~{Spurny}, Z.~{Ceplecha}, J.~{Rendtel}, H.~{Betlem}, Meteoritics and
  Planetary Science \textbf{33}(1), 49 (1998).
\newblock \doi{10.1111/j.1945-5100.1998.tb01606.x}

\bibitem{2011Icar..216...40J}
P.~{Jenniskens}, P.S. {Gural}, L.~{Dynneson}, B.J. {Grigsby}, K.E. {Newman},
  M.~{Borden}, M.~{Koop}, D.~{Holman}, Icarus \textbf{216}(1), 40 (2011).
\newblock \doi{10.1016/j.icarus.2011.08.012}

\bibitem{2012AuJES..59..177B}
P.A. {Bland}, P.~{Spurn{\'y}}, A.W.R. {Bevan}, K.T. {Howard}, M.C. {Towner},
  G.K. {Benedix}, R.C. {Greenwood}, L.~{Shrben{\'y}}, I.A. {Franchi},
  G.~{Deacon}, Australian Journal of Earth Sciences \textbf{59}(2), 177 (2012).
\newblock \doi{10.1080/08120099.2011.595428}

\bibitem{2015P&SS..118..102T}
J.~{T{\'o}th}, L.~{Korno{\v s}}, P.~{Zigo}, {\v S}.~{Gajdo{\v s}},
  D.~{Kalman{\v c}ok}, J.~{Vil{\'a}gi}, J.~{{\v S}imon}, P.~{Vere{\v s}},
  J.~{{\v S}ilha}, M.~{Bu{\v c}ek}, A.~{Gal{\'a}d}, P.~{Rus{\v n}{\'a}k},
  P.~{Hr{\'a}bek}, F.~{{\v D}uri{\v s}}, R.~{Rudawska}, Planetary and Space
  Science \textbf{118}, 102 (2015).
\newblock \doi{10.1016/j.pss.2015.07.007}

\bibitem{2015pimo.conf...37C}
F.~{Colas}, B.~{Zanda}, J.~{Vaubaillon}, S.~{Bouley}, C.~{Marmo},
  Y.~{Audureau}, M.K. {Kwon}, J.L. {Rault}, S.~{Caminade}, P.~{Vernazza}, in
  \emph{International Meteor Conference Mistelbach, Austria} (2015), p.~37

\bibitem{2010JIMO...38...54V}
P.~{Vere{\v{s}}}, J.~{Toth}, WGN, Journal of the International Meteor
  Organization \textbf{38}(2), 54 (2010)

\bibitem{2014me13.conf..225K}
L.~{Korno{\v s}}, P.~{Matlovi{\v c}}, R.~{Rudawska}, J.~{T{\'o}th},
  M.~{Hajdukov{\'a}}, Jr., J.~{Koukal}, R.~{Piffl}, Meteoroids 2013 pp.
  225--233 (2014)

\bibitem{1994QJRAS..35..293B}
W.J. {Baggaley}, R.G.T. {Bennett}, D.I. {Steel}, A.D. {Taylor}, Quarterly
  Journal of the Royal Astronomical Society \textbf{35}, 293 (1994)

\bibitem{2004EM&P...95..221W}
R.J. {Weryk}, P.~{Brown}, Earth Moon and Planets \textbf{95}(1-4), 221 (2004).
\newblock \doi{10.1007/s11038-005-9034-x}

\bibitem{2015ApJ...809...36J}
D.~{Janches}, S.~{Close}, J.L. {Hormaechea}, N.~{Swarnalingam}, A.~{Murphy},
  D.~{O'Connor}, B.~{Vand epeer}, B.~{Fuller}, D.C. {Fritts}, C.~{Brunini}, The
  Astrophysical Journal \textbf{809}(1), 36 (2015).
\newblock \doi{10.1088/0004-637X/809/1/36}

\bibitem{2016A&A...592A.150P}
P.~{Pokorn{\'y}}, P.G. {Brown}, Astronomy \& Astrophysics \textbf{592}, A150
  (2016).
\newblock \doi{10.1051/0004-6361/201628134}

\bibitem{2011MNRAS.414.3322B}
R.C. {Blaauw}, M.D. {Campbell-Brown}, R.J. {Weryk}, Monthly Notices of the
  Royal Astronomical Society \textbf{414}(4), 3322 (2011).
\newblock \doi{10.1111/j.1365-2966.2011.18633.x}

\bibitem{1985Icar...62..244G}
E.~{Grun}, H.A. {Zook}, H.~{Fechtig}, R.H. {Giese}, Icarus \textbf{62}(2), 244
  (1985).
\newblock \doi{10.1016/0019-1035(85)90121-6}

\bibitem{2006CoSka..36..103P}
V.~{Porub{\v c}an}, L.~{Korno{\v s}}, I.P. {Williams}, Contributions of the
  Astronomical Observatory Skalnate Pleso \textbf{36}, 103 (2006)

\bibitem{1993MNRAS.264...93A}
D.J. {Asher}, S.V.M. {Clube}, D.I. {Steel}, Monthly Notices of the Royal
  Astronomical Society \textbf{264}, 93 (1993).
\newblock \doi{10.1093/mnras/264.1.93}

\bibitem{2017A&A...605A..68S}
P.~{Spurn{\'y}}, J.~{Borovi{\v c}ka}, H.~{Mucke}, J.~{Svore{\v n}}, Astronomy
  \& Astrophysics \textbf{605}, A68 (2017).
\newblock \doi{10.1051/0004-6361/201730787}

\bibitem{2010AJ....139.1822J}
P.~{Jenniskens}, J.~{Vaubaillon}, Astronomical Journal \textbf{139}, 1822
  (2010).
\newblock \doi{10.1088/0004-6256/139/5/1822}

\bibitem{2017P&SS..143....3J}
T.J. {Jopek}, Z.~{Ka{\v n}uchov{\'a}}, Planetary and Space Science
  \textbf{143}, 3 (2017).
\newblock \doi{10.1016/j.pss.2016.11.003}

\bibitem{2019msme.book..210W}
I.P. {Williams}, T.J. {Jopek}, R.~{Rudawska}, J.~{T{\'o}th}, L.~{Korno{\v{s}}},
  \emph{{Minor Meteor Showers and the Sporadic Background}} (2019), p. 210

\bibitem{1963SCoA....7..261S}
R.B. {Southworth}, G.S. {Hawkins}, Smithsonian Contributions to Astrophysics
  \textbf{7}, 261 (1963)

\bibitem{1981Icar...45..545D}
J.D. {Drummond}, Icarus \textbf{45}(3), 545 (1981).
\newblock \doi{10.1016/0019-1035(81)90020-8}

\bibitem{1999MNRAS.304..743V}
G.B. {Valsecchi}, T.J. {Jopek}, C.~{Froeschle}, Monthly Notices of the Royal
  Astronomical Society \textbf{304}(4), 743 (1999).
\newblock \doi{10.1046/j.1365-8711.1999.02264.x}

\bibitem{2008EM&P..102...73J}
T.J. {Jopek}, R.~{Rudawska}, P.~{Bartczak}, Earth Moon and Planets
  \textbf{102}(1-4), 73 (2008).
\newblock \doi{10.1007/s11038-007-9197-8}

\bibitem{2016MNRAS.455.4329M}
A.V. {Moorhead}, Monthly Notices of the Royal Astronomical Society
  \textbf{455}(4), 4329 (2016).
\newblock \doi{10.1093/mnras/stv2610}

\bibitem{2017A&A...598A..40N}
L.~{Neslu{\v{s}}an}, M.~{Hajdukov{\'a}}, Astronomy \& Astrophysics
  \textbf{598}, A40 (2017).
\newblock \doi{10.1051/0004-6361/201629659}

\bibitem{2015P&SS..118...38R}
R.~{Rudawska}, P.~{Matlovi{\v{c}}}, J.~{T{\'o}th}, L.~{Korno{\v{s}}}, Planetary
  and Space Science \textbf{118}, 38 (2015).
\newblock \doi{10.1016/j.pss.2015.07.011}

\bibitem{2008Icar..195..317B}
P.~{Brown}, R.J. {Weryk}, D.K. {Wong}, J.~{Jones}, Icarus \textbf{195}(1), 317
  (2008).
\newblock \doi{10.1016/j.icarus.2007.12.002}

\bibitem{2005A&A...439..761V}
J.~{Vaubaillon}, F.~{Colas}, L.~{Jorda}, Astronomy \& Astrophysics
  \textbf{439}(2), 761 (2005).
\newblock \doi{10.1051/0004-6361:20042626}

\bibitem{2013SoSyR..47..219R}
G.O. {Ryabova}, Solar System Research \textbf{47}(3), 219 (2013).
\newblock \doi{10.1134/S0038094613030052}

\bibitem{2005MNRAS.359..551G}
D.P. {Galligan}, W.J. {Baggaley}, Monthly Notices of the Royal Astronomical
  Society \textbf{359}(2), 551 (2005).
\newblock \doi{10.1111/j.1365-2966.2005.08918.x}

\bibitem{2017M&PS...52.1683B}
A.~{Bischoff}, J.A. {Barrat}, K.~{Bauer}, C.~{Burkhardt}, H.~{Busemann},
  S.~{Ebert}, M.~{Gonsior}, J.~{Hakenm{\"u}ller}, J.~{Haloda}, D.~{Harries},
  Meteoritics and Planetary Science \textbf{52}(8), 1683 (2017).
\newblock \doi{10.1111/maps.12883}

\bibitem{1993MNRAS.265..524J}
J.~{Jones}, P.~{Brown}, Monthly Notices of the Royal Astronomical Society
  \textbf{265}, 524 (1993).
\newblock \doi{10.1093/mnras/265.3.524}

\bibitem{2010ApJ...713..816N}
D.~{Nesvorn{\'y}}, P.~{Jenniskens}, H.F. {Levison}, W.F. {Bottke},
  D.~{Vokrouhlick{\'y}}, M.~{Gounelle}, The Astrophysical Journal
  \textbf{713}(2), 816 (2010).
\newblock \doi{10.1088/0004-637X/713/2/816}

\bibitem{2014ApJ...789...25P}
P.~{Pokorn{\'y}}, D.~{Vokrouhlick{\'y}}, D.~{Nesvorn{\'y}},
  M.~{Campbell-Brown}, P.~{Brown}, The Astrophysical Journal \textbf{789}(1),
  25 (2014).
\newblock \doi{10.1088/0004-637X/789/1/25}

\bibitem{2008Icar..196..144C}
M.D. {Campbell-Brown}, Icarus \textbf{196}(1), 144 (2008).
\newblock \doi{10.1016/j.icarus.2008.02.022}

\bibitem{1999P&SS...47.1005M}
N.~{McBride}, J.a.m. {McDonnell}, Planetary and Space Science \textbf{47}(8-9),
  1005 (1999).
\newblock \doi{10.1016/S0032-0633(99)00023-9}

\bibitem{2017P&SS..143..209M}
A.V. {Moorhead}, P.G. {Brown}, M.D. {Campbell-Brown}, D.~{Heynen}, W.J.
  {Cooke}, Planetary and Space Science \textbf{143}, 209 (2017).
\newblock \doi{10.1016/j.pss.2017.02.002}

\bibitem{1993Natur.362..428G}
E.~{Grun}, H.A. {Zook}, M.~{Baguhl}, A.~{Balogh}, S.J. {Bame}, H.~{Fechtig},
  R.~{Forsyth}, M.S. {Hanner}, M.~{Horanyi}, J.~{Kissel}, Nature
  \textbf{362}(6419), 428 (1993).
\newblock \doi{10.1038/362428a0}

\bibitem{1996Natur.380..323T}
A.D. {Taylor}, W.J. {Baggaley}, D.I. {Steel}, Nature \textbf{380}(6572), 323
  (1996).
\newblock \doi{10.1038/380323a0}

\bibitem{2014M&PS...49...63H}
M.~{Hajdukov{\'a}}, L.~{Korno{\v{s}}}, J.~{T{\'o}th}, Meteoritics and Planetary
  Science \textbf{49}(1), 63 (2014).
\newblock \doi{10.1111/maps.12119}

\bibitem{2007AdSpR..39..491J}
P.~{Jenniskens}, Advances in Space Research \textbf{39}, 491 (2007).
\newblock \doi{10.1016/j.asr.2007.03.040}

\bibitem{2017P&SS..143..104M}
P.~{Matlovi{\v c}}, J.~{T{\'o}th}, R.~{Rudawska}, L.~{Korno{\v s}}, Planetary
  and Space Science \textbf{143}, 104 (2017).
\newblock \doi{10.1016/j.pss.2017.02.007}

\bibitem{2014EM&P..112...45R}
R.~{Rudawska}, J.~{Zender}, P.~{Jenniskens}, J.~{Vaubaillon}, P.~{Koten},
  A.~{Margonis}, J.~{T{\'o}th}, J.~{McAuliffe}, D.~{Koschny}, Earth Moon and
  Planets \textbf{112}, 45 (2014).
\newblock \doi{10.1007/s11038-014-9436-8}

\bibitem{2017P&SS..143..238M}
J.M. {Madiedo}, Planetary and Space Science \textbf{143}, 238 (2017).
\newblock \doi{10.1016/j.pss.2016.12.005}

\bibitem{2019A&A...621A..68V}
V.~{Voj{\'a}{\v c}ek}, J.~{Borovi{\v c}ka}, P.~{Koten}, P.~{Spurn{\'y}},
  R.~{{\v S}tork}, Astronomy \& Astrophysics \textbf{621}, A68 (2019).
\newblock \doi{10.1051/0004-6361/201833289}

\bibitem{2016Icar..278..248B}
J.~{Borovi{\v c}ka}, A.A. {Berezhnoy}, Icarus \textbf{278}, 248 (2016).
\newblock \doi{10.1016/j.icarus.2016.06.022}

\bibitem{2004AsBio...4...67J}
P.~{Jenniskens}, E.L. {Schaller}, C.O. {Laux}, M.A. {Wilson}, G.~{Schmidt},
  R.L. {Rairden}, Astrobiology \textbf{4}(1), 67 (2004).
\newblock \doi{10.1089/153110704773600249}

\bibitem{2010Icar..210..150B}
A.A. {Berezhnoy}, J.~{Borovi{\v{c}}ka}, Icarus \textbf{210}(1), 150 (2010).
\newblock \doi{10.1016/j.icarus.2010.06.036}

\bibitem{2004AsBio...4..123J}
P.~{Jenniskens}, A.M. {Mandell}, Astrobiology \textbf{4}(1), 123 (2004).
\newblock \doi{10.1089/153110704773600285}

\bibitem{2012JGRA..117.9323C}
M.D. {Campbell-Brown}, J.~{Kero}, C.~{Szasz}, A.~{Pellinen-Wannberg}, R.J.
  {Weryk}, Journal of Geophysical Research (Space Physics) \textbf{117}(A9),
  A09323 (2012).
\newblock \doi{10.1029/2012JA017800}

\bibitem{2011Icar..212..877G}
M.~{Gritsevich}, D.~{Koschny}, Icarus \textbf{212}(2), 877 (2011).
\newblock \doi{10.1016/j.icarus.2011.01.033}

\bibitem{1988BAICz..39..221C}
Z.~{Ceplecha}, Bulletin of the Astronomical Institutes of Czechoslovakia
  \textbf{39}, 221 (1988)

\bibitem{2009A&A...495..353B}
P.B. {Babadzhanov}, G.I. {Kokhirova}, Astronomy \& Astrophysics \textbf{495},
  353 (2009).
\newblock \doi{10.1051/0004-6361:200810460}

\bibitem{2018DPS....5010001V}
D.~{Vida}, P.~{Brown}, M.~{Campbell-Brown}, in \emph{AAS/Division for Planetary
  Sciences Meeting Abstracts \#50} (2018), AAS/Division for Planetary Sciences
  Meeting Abstracts, p. 100.01

\bibitem{2019A&A...625A.106C}
D.~{{\v{C}}apek}, P.~{Koten}, J.~{Borovi{\v{c}}ka}, V.~{Voj{\'a}{\v{c}}ek},
  P.~{Spurn{\'y}}, R.~{{\v{S}}tork}, Astronomy \& Astrophysics \textbf{625},
  A106 (2019).
\newblock \doi{10.1051/0004-6361/201935203}

\bibitem{2013A&A...557A..41C}
M.D. {Campbell-Brown}, J.~{Borovi{\v{c}}ka}, P.G. {Brown}, E.~{Stokan},
  Astronomy \& Astrophysics \textbf{557}, A41 (2013).
\newblock \doi{10.1051/0004-6361/201322005}

\bibitem{2015MNRAS.447.1580S}
E.~{Stokan}, M.D. {Campbell-Brown}, Monthly Notices of the Royal Astronomical
  Society \textbf{447}(2), 1580 (2015).
\newblock \doi{10.1093/mnras/stu2552}

\end{thebibliography}

\end{document}